\documentclass[
reprint,
superscriptaddress,
amsmath,amssymb,
aps,
prb,
]{revtex4-2}

\usepackage{graphicx}
\usepackage{dcolumn}
\usepackage{bm}
\usepackage{xcolor}
\usepackage{hyperref}
\usepackage{placeins}

\begin{document}

\title{Boundary conditions dictate frequency dependence of thermal conductivity in silicon}

\author{Yizhe Liu}
\affiliation{Tsinghua SIGS, Tsinghua University, Shenzhen 518055, China}
\author{Qinshu Li}
\affiliation{Tsinghua-Berkeley Shenzhen Institute, Tsinghua University, Shenzhen 518055, China}
\author{Fang Liu}
\affiliation{State Key Laboratory for Mesoscopic Physics and Frontiers Science Center for Nano-optoelectronics, School of Physics, Peking University, Beijing 100871, China}
\author{Xinqiang Wang}
\affiliation{State Key Laboratory for Mesoscopic Physics and Frontiers Science Center for Nano-optoelectronics, School of Physics, Peking University, Beijing 100871, China}
\affiliation{Peking University Yangtze Delta Institute of Optoelectronics, Nantong, Jiangsu 226010, China}
\author{Bo Sun}
\email{sun.bo@sz.tsinghua.edu.cn}
\affiliation{Tsinghua-Berkeley Shenzhen Institute, Tsinghua University, Shenzhen 518055, China}
\affiliation{Tsinghua Shenzhen International Graduate School and Guangdong Provincial Key Laboratory of Thermal Management Engineering $\&$ Materials, Shenzhen 518055, China}

\begin{abstract}
\noindent 
Non-Fourier thermal transports have drawn significant attention for decades. Among them, the frequency dependent thermal conductivity has been extensively explored by pump-probe techniques, such as time-domain thermoreflectance, which is employed to probe the spectra of phonon mean free paths. However, previous studies on silicon have not exhibited apparent frequency dependence despite its broad phonon distribution. Here, we report the frequency dependent thermal transport in Al/Si with an atomically sharp interface, where the matched Debye temperatures preserve non-equilibrium between low- and high-energy phonons in Si. The dependence vanishes in Al/SiO$_2$/Si at room temperature, since the SiO$_2$ interlayer facilitates phonon scattering and destroys thermal non-equilibrium. At 80 K, frequency dependence reemerges in Al/SiO$_2$/Si, due to reduced interfacial phonon scattering. Our findings highlight the significance of boundary conditions in frequency dependent thermal conductivity. 
\end{abstract}

\maketitle

\section{Introduction}
\label{sec: Introduction}

Heat transports diffusively when carrier mean free paths (MFPs) are shorter than the characteristic length of the temperature profile, as described by Fourier's law. In high-frequency electronic devices, periodic Joule heating leads to a restricted temperature profile \cite{1}. As a result, the important temperature lengths diminish to the carrier MFPs, and non-Fourier transport appears \cite{2,3,4,5,6,7,8}. Consequently, thermal conductivity ($\Lambda$) shows a dependence on heating frequency, with a marked suppression when the heat diffusion length approaches phonon MFPs \cite{9}.

Experimental demonstrations on the frequency dependence of thermal transport have been achieved through pump-probe studies with periodic heating at several MHz on semiconductor alloys \cite{10}, and later on graphite \cite{11} and other two-dimensional materials \cite{12,13}. The observed frequency dependent $\Lambda$ was initially attributed to a broad spectrum of phonon MFPs by Koh \textit{et al.} \cite{10}, where phonons with MFPs longer than the important temperature lengths could not be probed, thus motivating innovative experiments to measure phonon spectra. \cite{11,12,14,15,16}. Furthermore, Wilson \textit{et al.} interpreted the frequency dependent thermal transport using a two-channel model based on the phonon non-equilibrium, where weak interactions between low- and high-frequency phonons constitutes additional non-equilibrium resistance \cite{17}, dictated by the transducer/sample interface. This model successfully explained the reduction in apparent $\Lambda$ (at high frequencies) and thermal conductance \textit{G} (at low frequencies) in SiGe alloys \cite{17} and transition metal dichalcogenides \cite{13}, enriching the understanding of phonon dynamics. Additionally, a ballistic/diffusive model was proposed to explain non-Fourier transports within time-domain thermoreflectance (TDTR) and broadband frequency-domain thermoreflectance (BB-FDTR) experiments, where ballistic phonons carry less heat than Fourier’s law predictions and the reduced $\Lambda$ originates from the reflection of long MFP phonons at the interface \cite{18}. Importantly, both theories reveal the critical influence of the interface on frequency dependent thermal transport, thus underscoring the significance of the transducer/sample boundary condition. 

However, a comprehensive understanding of the observed frequency dependence remains elusive with certain puzzles. For example, Si displays a broader spectral distribution of phonon MFPs, compared to two-dimensional materials displaying frequency dependent $\Lambda$ along the through-plane direction \cite{14,19}. Nevertheless, previous TDTR measurements on Si at room temperature (RT) did not reveal apparent frequency dependence, regardless of the transducer/Si boundary conditions. Moreover, although previous experiments have reported that the transducer/sample interface affects the frequency dependence \cite{14,15,18}, the underlying mechanisms is still incomplete, impeding the accurate measurements of intrinsic thermal properties.

In this work, we discover a pronounced frequency dependent thermal transport in TDTR measurements on Al/Si with an atomically sharp interface. The atomically sharp Al/Si interface, with matched phonon density of states (DOS), minimizes the interfacial phonon scattering, preserving the phonon non-equilibrium in Si. In contrast, the Al/SiO$_{2}$/Si interface with a SiO$_{2}$ barrier layer effectively scatters phonons and facilitates energy exchange within different branches of phonons, preventing frequency dependence at RT.  At 80 K, the Al/SiO$_2$/Si sample exhibited a resurgence of frequency dependent thermal transport, resulting from the reduced phonon scattering at the interface. Furthermore, our measurements on the Al/Si sample successfully extract the spectrum of phonon MFPs in Si at RT and 80 K, in agreement with first-principles calculations. Our observations emphasize the vital role of boundary conditions in frequency dependent thermal transport, which has significant implications for the design and thermal management of high-frequency electronic devices.Heat transports diffusively when carrier mean free paths (MFPs) are shorter than the characteristic length of the temperature profile, as described by Fourier's law. In high-frequency electronic devices, periodic Joule heating leads to a restricted temperature profile \cite{1}. As a result, the important temperature lengths diminish to the carrier MFPs, and non-Fourier transport appears \cite{2,3,4,5,6,7,8}. Consequently, thermal conductivity ($\Lambda$) shows a dependence on heating frequency, with a marked suppression when the heat diffusion length approaches phonon MFPs \cite{9}.

Experimental demonstrations on the frequency dependence of thermal transport have been achieved through pump-probe studies with periodic heating at several MHz on semiconductor alloys \cite{10}, and later on graphite \cite{11} and other two-dimensional materials \cite{12,13}. The observed frequency dependent $\Lambda$ was initially attributed to a broad spectrum of phonon MFPs by Koh \textit{et al.} \cite{10}, where phonons with MFPs longer than the important temperature lengths could not be probed, thus motivating innovative experiments to measure phonon spectra. \cite{11,12,14,15,16}. Furthermore, Wilson \textit{et al.} interpreted the frequency dependent thermal transport using a two-channel model based on the phonon non-equilibrium, where weak interactions between low- and high-frequency phonons constitutes additional non-equilibrium resistance \cite{17}, dictated by the transducer/sample interface. This model successfully explained the reduction in apparent $\Lambda$ (at high frequencies) and thermal conductance \textit{G} (at low frequencies) in SiGe alloys \cite{17} and transition metal dichalcogenides \cite{13}, enriching the understanding of phonon dynamics. Additionally, a ballistic/diffusive model was proposed to explain non-Fourier transports within time-domain thermoreflectance (TDTR) and broadband frequency-domain thermoreflectance (BB-FDTR) experiments, where ballistic phonons carry less heat than Fourier’s law predictions and the reduced $\Lambda$ originates from the reflection of long MFP phonons at the interface \cite{18}. Importantly, both theories reveal the critical influence of the interface on frequency dependent thermal transport, thus underscoring the significance of the transducer/sample boundary condition. 

However, a comprehensive understanding of the observed frequency dependence remains elusive with certain puzzles. For example, Si displays a broader spectral distribution of phonon MFPs, compared to two-dimensional materials displaying frequency dependent $\Lambda$ along the through-plane direction \cite{14,19}. Nevertheless, previous TDTR measurements on Si at room temperature (RT) did not reveal apparent frequency dependence, regardless of the transducer/Si boundary conditions. Moreover, although previous experiments have reported that the transducer/sample interface affects the frequency dependence \cite{14,15,18}, the underlying mechanisms is still incomplete, impeding the accurate measurements of intrinsic thermal properties.

In this work, we discover a pronounced frequency dependent thermal transport in TDTR measurements on Al/Si with an atomically sharp interface. The atomically sharp Al/Si interface, with matched phonon density of states (DOS), minimizes the interfacial phonon scattering, preserving the phonon non-equilibrium in Si. In contrast, the Al/SiO$_{2}$/Si interface with a SiO$_{2}$ barrier layer effectively scatters phonons and facilitates energy exchange within different branches of phonons, preventing frequency dependence at RT.  At 80 K, the Al/SiO$_2$/Si sample exhibited a resurgence of frequency dependent thermal transport, resulting from the reduced phonon scattering at the interface. Furthermore, our measurements on the Al/Si sample successfully extract the spectrum of phonon MFPs in Si at RT and 80 K, in agreement with first-principles calculations. Our observations emphasize the vital role of boundary conditions in frequency dependent thermal transport, which has significant implications for the design and thermal management of high-frequency electronic devices.

\section{Experimental details}
\label{sec: Experimental details}
To understand the impact of boundary conditions on the frequency dependent thermal transport, we designed two samples with different boundary conditions. For Boundary 1, we fabricated a high-quality Al/Si interface with atomic sharpness through molecular beam epitaxy growth of Al (111) on Si (111), as detailed in our prior works \cite{20,21}. According to the diffuse mismatch model (DMM), the similar Debye temperatures with matched phonon DOS provide efficient phonon transmission across the interface, leading to a low phonon scattering rate. Consequently, phonon transport across this interface is predominantly diffusive, with preserved thermal non-equilibrium between low- and high-energy phonons. For comparison, we built an interface with a much higher rate of phonon scattering to facilitate sufficient energy exchange within different branches of phonons (Boundary 2). Due to the difficulty in growing metals on Si with highly dissimilar Debye temperatures while maintaining an atomically sharp interface, we fabricated an Al/SiO$_2$/Si sample instead to enhance phonon scattering at the interface. Using magnetron sputtering deposition, we grew a thin Al layer on the Si substrate without moving its native oxide. The Al layer thickness in both samples was measured by picosecond acoustics, 138 nm for the Al/Si sample and 70 nm for the Al/SiO$_2$/Si sample, respectively.

We measured $\Lambda$ of Si and \textit{G} of the Al/Si interface in as-prepared samples by TDTR \cite{22,23}, a non-contact pump-probe technique to characterize thermal transport in a variety of materials \cite{37,38,39}. Briefly, we split the pulsed femtosecond laser beam (785 nm, 80 MHz repetition rate) into a pump beam and a probe beam using a polarizing beam splitter, with the relative delay time adjusted by a mechanical linear motor stage. We modulated the pump laser beam at a radio frequency \textit{f}, ranging from 0.47 MHz to 10.1 MHz through an electro-optical modulator (EOM) to induce periodic temperature oscillation on the sample surface. The probe beam monitored the corresponding temperature-dependent reflectance changes of the sample surface, modulated at an audio frequency of 200 Hz by a mechanical chopper to enhance the signal-to-noise ratio. We then demodulated the acquired signals with lock-in amplifiers. The pump and probe beams were focused to a 1/\textit{e}$^{2}$ radius of 30 $\mu$m to minimize the non-equilibrium, induced by the spot size in our measurements \cite{14}. 

We initially fitted the demodulated signals (ratio of the in-phase component and the out-of-phase component of the lock-in amplifier) with a stratified thermal model for extracting thermal properties \cite{22,24}. The model utilized a numerical solution of the heat diffusion equation, grounded in the local-equilibrium assumption, where all energy carriers share the same temperature. This assumption implies, for instance, that there are no distinct temperature gradients between different phonon branches within the material. Accordingly, a single thermal transport channel with specified thermal parameters is sufficient for our fitting, denoted as the one-channel model with further details provided in Supplemental Material \cite{25}. Frequency dependent thermal transport was only observed within the Al/Si sample at RT, while it came forth in both samples at 80 K. We then interpreted observed frequency dependence by the two-channel model, considering non-equilibrium heat flow between low- and high-energy phonons in our TDTR measurements (see Supplemental Material \cite{25}).
\begin{figure}
\includegraphics[width=0.48\textwidth,height=0.34\textwidth]{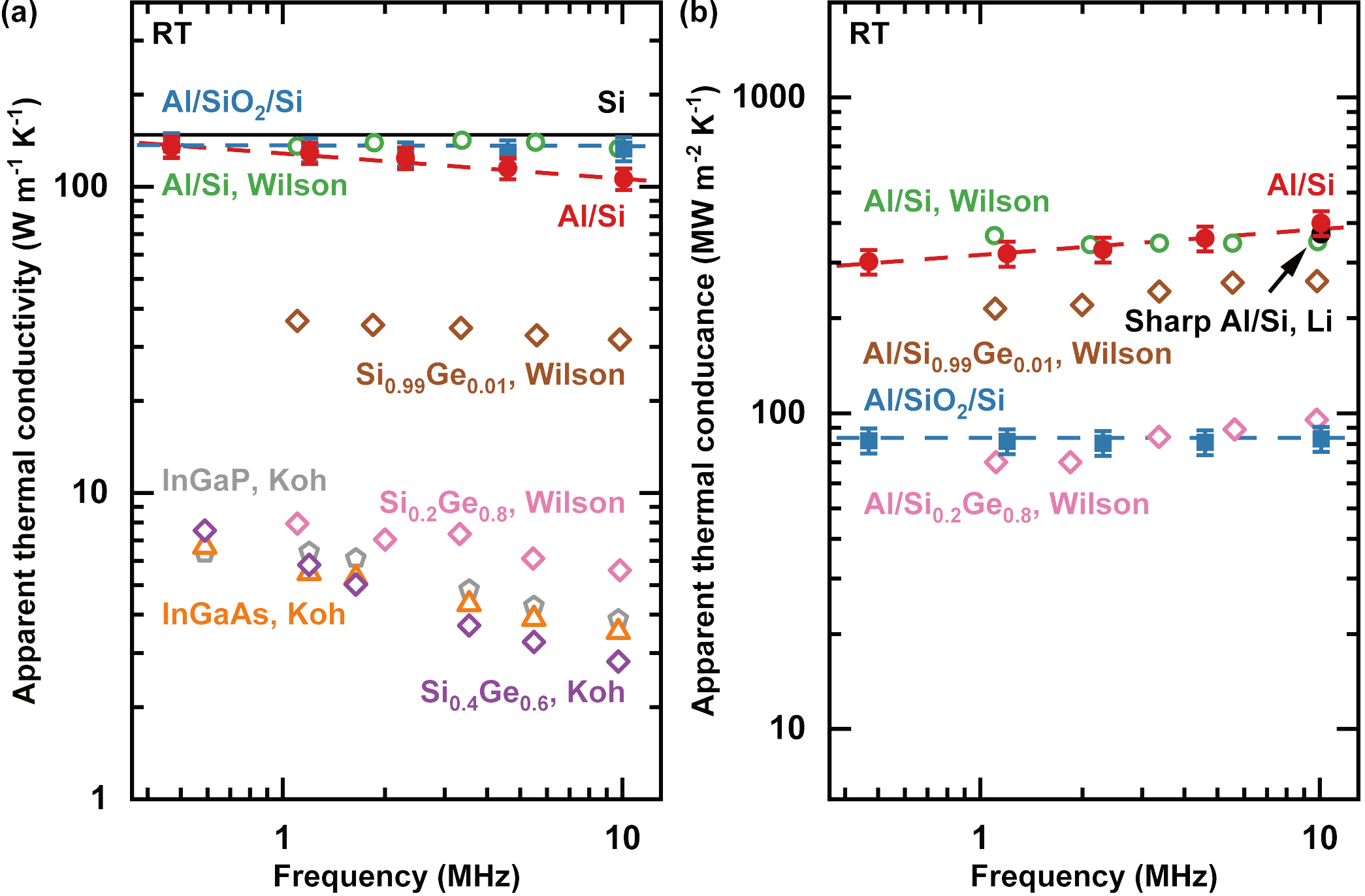}%
\caption{\label{fig_1}Apparent (a) $\Lambda$ and (b) \textit{G} of Al/Si (solid red circles) and Al/SiO$_2$/Si (solid blue squares) samples, derived from the one-channel thermal model at RT. The solid black line represents $\Lambda$ of bulk Si \cite{26} while the solid black circle stands for \textit{G} of sharp Al/Si interface by Li \textit{et al} \cite{20}. Open diamonds exhibit reported frequency dependence in Si$_{0.99}$Ge$_{0.01}$ (brown) \cite{17}, Si$_{0.2}$Ge$_{0.8}$ (pink) \cite{18}, and Si$_{0.4}$Ge$_{0.6}$ (purple) \cite{11}. Orange triangles and grey pentagons show the frequency-dependenct $\Lambda$ measured in In$_{0.53}$Ga$_{0.47}$As and In$_{0.49}$Ga$_{0.51}$P, respectively \cite{11}. }
\end{figure}

\section{Results and discussion}
\subsection{Frequency dependence at room temperature}
\label{subsec: Frequency dependence at room temperature}

We first conducted TDTR measurements at RT. Apparent $\Lambda$ of Si and \textit{G} of the Al/Si interface were extracted by the one-channel model, as depicted in Fig. 1. In the Al/Si sample with Boundary 1, $\Lambda$ exhibits a pronounced frequency dependence, decreasing from 135 W m$^{-1}$ K$^{-1}$ at 0.47 MHz to 106 W m$^{-1}$ K$^{-1}$ at 10.1 MHz, which has not been observed before in Al/Si sample even with a clean interface, where native oxide is removed. We compare our results with previously reported $\Lambda$ in Al/semiconductor alloys displaying frequency dependence, including Si$_{0.99}$Ge$_{0.01}$ \cite{17}, Si$_{0.2}$Ge$_{0.8}$ \cite{18}, Si$_{0.4}$Ge$_{0.6}$ \cite{20}, In$_{0.53}$Ga$_{0.47}$As \cite{20}, and In$_{0.49}$Ga$_{0.51}$P \cite{20}. We find that $\Lambda$ of the Al/Si sample exhibits a comparable decreasing trend with \textit{f} to that observed in Si$_{0.99}$Ge$_{0.01}$. Meanwhile, \textit{G} increases with rising \textit{f}, reaching a maximum value of 385 MW m$^{-2}$ K$^{-1}$ at 10.1 MHz and a minimum value of 302 MW m$^{-2}$ K$^{-1}$ at 0.47 MHz. The obtained maximum values of $\Lambda$ and \textit{G} in the Al/Si sample are in excellent agreement with literature values within the error range \cite{20,26}, validating the accuracy of our measurements. Then we plot representative TDTR signals with corresponding fitting curves from the one-channel model, presented in Fig. 2(a). Applying one set of parameters ($\Lambda$=135 W m$^{-1}$ K$^{-1}$ and \textit{G}=302 MW m$^{-2}$ K$^{-1}$), the one-channel model fits well for data at 0.47 MHz but shows an obvious overestimation for data at higher frequencies, exceeding the error range. This discrepancy reveals a suppressed $\Lambda$ under high-frequency periodic heating. Notably, although earlier TDTR measurements on Al/Si structures reported no discernible dependence with \textit{f} at RT, the frequency dependence appears in our Al/Si sample, indicating a crucial role for the high-quality interface with atomic-level sharpness. As documented in our prior studies, the transmissivity of low-energy phonons considerably exceeds that of high-energy phonons, thus leading to a substantial non-equilibrium thermal resistance \cite{20}. At low modulation frequencies, the additional resistance manifests as a notable reduction in \textit{G}. Conversely, at higher modulation frequencies, TDTR signals are more sensitive to thermal properties near the interface, resulting in the non-equilibrium resistance perceived as a decrease in $\Lambda$. The atomically sharp interface in our Al/Si sample preserves the non-equilibrium effects and reveals the frequency dependence in our TDTR measurements.
\begin{figure}
\includegraphics[width=0.36\textwidth,height=0.54\textwidth]{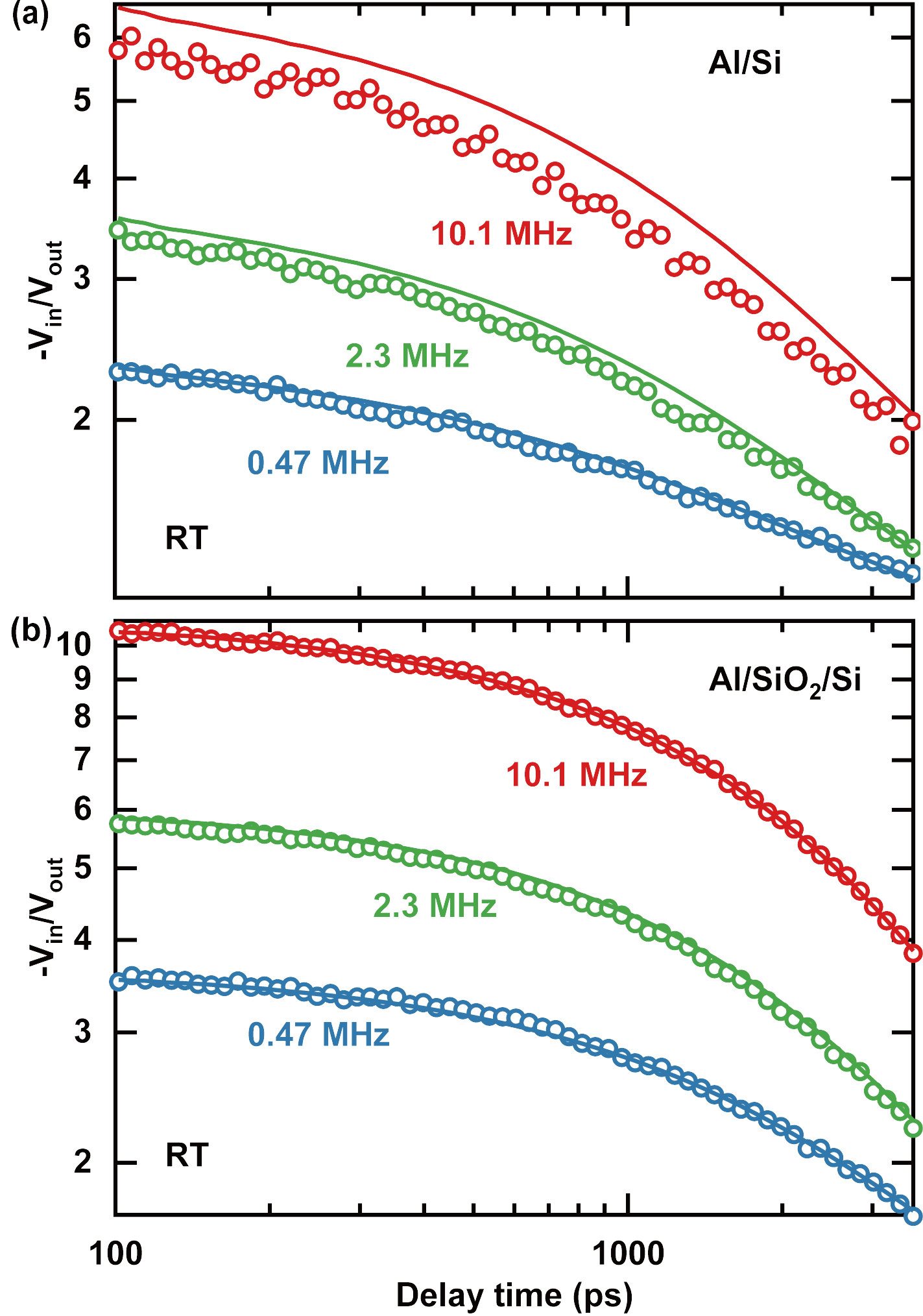}%
\caption{\label{fig_2}Fits of the one-channel thermal model (solid lines) to TDTR signals (open circles) measured in (a) the Al/Si sample and (b) the Al/SiO$_2$/Si sample at various modulation frequencies. Red, green, and blue plots represent signals at frequencies of 10.1 MHz, 2.3 MHz, and 0.47 MHz, respectively. In these fits, $\Lambda$ of Si is fixed at 134 W m$^{-1}$ K$^{-1}$ for all six curves. \textit{G} is set as 302 MW m$^{-2}$ K$^{-1}$ for the Al/Si sample and 81 MW m$^{-2}$ K$^{-1}$ for the Al/SiO$_2$/Si sample.}
\end{figure}

To further elucidate the impact of the boundary condition, we conducted TDTR experiments on the Al/SiO$_2$/Si sample with Boundary 2 at RT, where the presence of a native oxide layer between Al and Si constitutes a diffuse interface. In this sample, the measured $\Lambda$ ranges from 129 W m$^{-1}$ K$^{-1}$ to 136 W m$^{-1}$ K$^{-1}$ at varying frequencies, while \textit{G} varies from 80.1 MW m$^{-2}$ K$^{-1}$ to 82.2 MW m$^{-2}$ K$^{-1}$, demonstrating no apparent frequency dependence within experimental uncertainty. Hence, a single parameter set ($\Lambda$=134 W m$^{-1}$ K$^{-1}$ and \textit{G} is sufficient to fit the entire frequency range of the acquired TDTR data, as shown in Fig. 2(b). The absence of frequency dependence in the Al/SiO$_2$/Si sample agrees with prior TDTR studies on Al/Si structures, suggesting that a diffuse thermal interface inhibits non-equilibrium heat flow in high-frequency TDTR measurements, thereby preventing the emergence of frequency dependence \cite{20}. As illustrated in Fig. 3, the diffuse interface introduces additional sources of phonon scattering, such as atomic intermixing, roughness, oxide interlayer, and contamination, facilitating sufficient energy exchange between different phonon branches and consequently significantly suppressing phonon non-equilibrium. 

Our findings violate the prevailing assumption of the thermal interface's role in the ballistic/diffusive model proposed by Wilson \textit{et al.}, which assumes that a low-quality interface would increase the population of reflected phonons and enhance the frequency dependence \cite{18}. In their TDTR measurements at RT, we notice that the apparent $\Lambda$ of Si shows no apparent frequency dependence in Al/Si, Ta/Si, and Al/11 nm SiO$_{2}$/Si samples for all frequencies, with only Al/11 nm SiO$_{2}$/Si exhibiting a slight decrease at 9.8 MHz. This decrease is possibly due to the reduced sensitivity for $\Lambda$ of Si at 9.8 MHz since the 11 nm-thick SiO$_{2}$ significantly increases the interfacial thermal resistance. It is worth noting that \textit{G} of the Al/Si interface reported by Wilson \textit{et al.} lies in the range of our measurements \cite{18}, implying a good interface with high \textit{G}. However, their lack of observed frequency dependence at RT can be attributed to the absence of atomic-level structural details regarding interface roughness and atomic intermixing, limiting the understanding of frequency dependence in TDTR measurements. Hence, the value of \textit{G} is not a decisive factor in mediating frequency dependent thermal transport, while a sharp interface with appropriate phonon spectra of constituent materials consists of the proper boundary conditions.
\begin{figure}
\includegraphics[width=0.48\textwidth,height=0.21\textwidth]{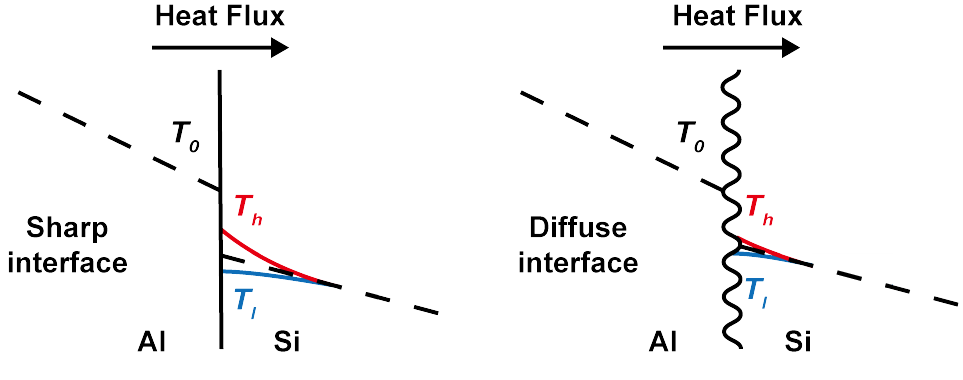}%
\caption{\label{fig_3}Schematic illustrating temperature distributions near the sharp and diffuse interfaces. Here \textit{T$_{0}$} represents temperature in Al near the interface. \textit{T$_{h}$} and \textit{T$_{l}$} are temperatures of high- and low-energy phonons in Si near the interface, respectively.}
\end{figure}

\subsection{Frequency dependence at 80 K}
\label{subsec: Frequency dependence at 80 K}

For a deeper insight into the influence of the boundary condition, we then conducted TDTR measurements on the two samples at 80 K, with the results exhibited in Fig. 4.  In the Al/Si sample with Boundary 1, we still observed the frequency dependence, with a relatively larger variation of 26.2 $\%$ in $\Lambda$, compared to the RT values. This is understandable since the phonon-phonon scattering is suppressed as the temperature decreases, resulting in an increased non-equilibrium thermal resistance under high heating frequencies. Importantly, despite the presence of a SiO$_2$ interlayer in the Al/SiO$_2$/Si sample, frequency dependence was observed at 80 K, with a change of 18.1 $\%$ in $\Lambda$, in sharp contrast to the room temperature behavior. The measured maximum $\Lambda$ in the Al/Si sample (745 W m$^{-1}$ K$^{-1}$) is lower than that in the Al/SiO$_2$/Si sample (890 W m$^{-1}$ K$^{-1}$) while both were smaller than the reference value measured using the steady-state method. This suggests a greater non-equilibrium thermal resistance in the Al/Si sample, owing to the sharp interface. Notably, we noted that both values were smaller than that reported in bulk Si. Previous studies have pointed that the measured inplane thermal conductivity $\Lambda$$_{//}$ shows a pronounced reduction when the temperature profile length falls below the phonon MFPs, leading to spot-size dependence in TDTR measurements \cite{14,18,27}. At 80 K, the longest phonon MFPs in Si reach $\sim$100 $\mu$m, exceeding the characteristic length of the temperature profile in our TDTR measurements \cite{14,28}. However, the 1/e$^{2}$ beam radius of 30 $\mu$m in our measurement results in a negligible sensitivity of signals to $\Lambda$$_{//}$ at \textit{f} higher than 4 MHz (see Supplemental Material \cite{25}). Consequently, the reduced apparent $\Lambda$ at 80 K comes from non-equilibrium thermal resistance due to high-frequency heating, rather than a decreased $\Lambda$$_{//}$ from finite laser spot size. 

The re-emerged frequency dependence in the Al/SiO$_{2}$/Si sample is attributed to two factors. Firstly, the strength of interactions between different phonon branches weakens at lower temperatures, resulting in inefficient energy transfer \cite{29}. A stronger phonon non-equilibrium facilitates the emergence of frequency dependence. Secondly, the phonon MFPs at 80 K are approximately an order of magnitude longer than those at RT \cite{14,19,28}. Therefore, the interface, even with a SiO$_{2}$ layer, is insufficient to scatter heat-carrying phonons, and the boundary condition transforms from Boundary 2 to Boundary 1 for the low-temperature case. This change leads to a pronounced temperature difference between low- and high-energy phonons and consequently observable non-equilibrium thermal resistance.
\begin{figure}
\includegraphics[width=0.48\textwidth,height=0.34\textwidth]{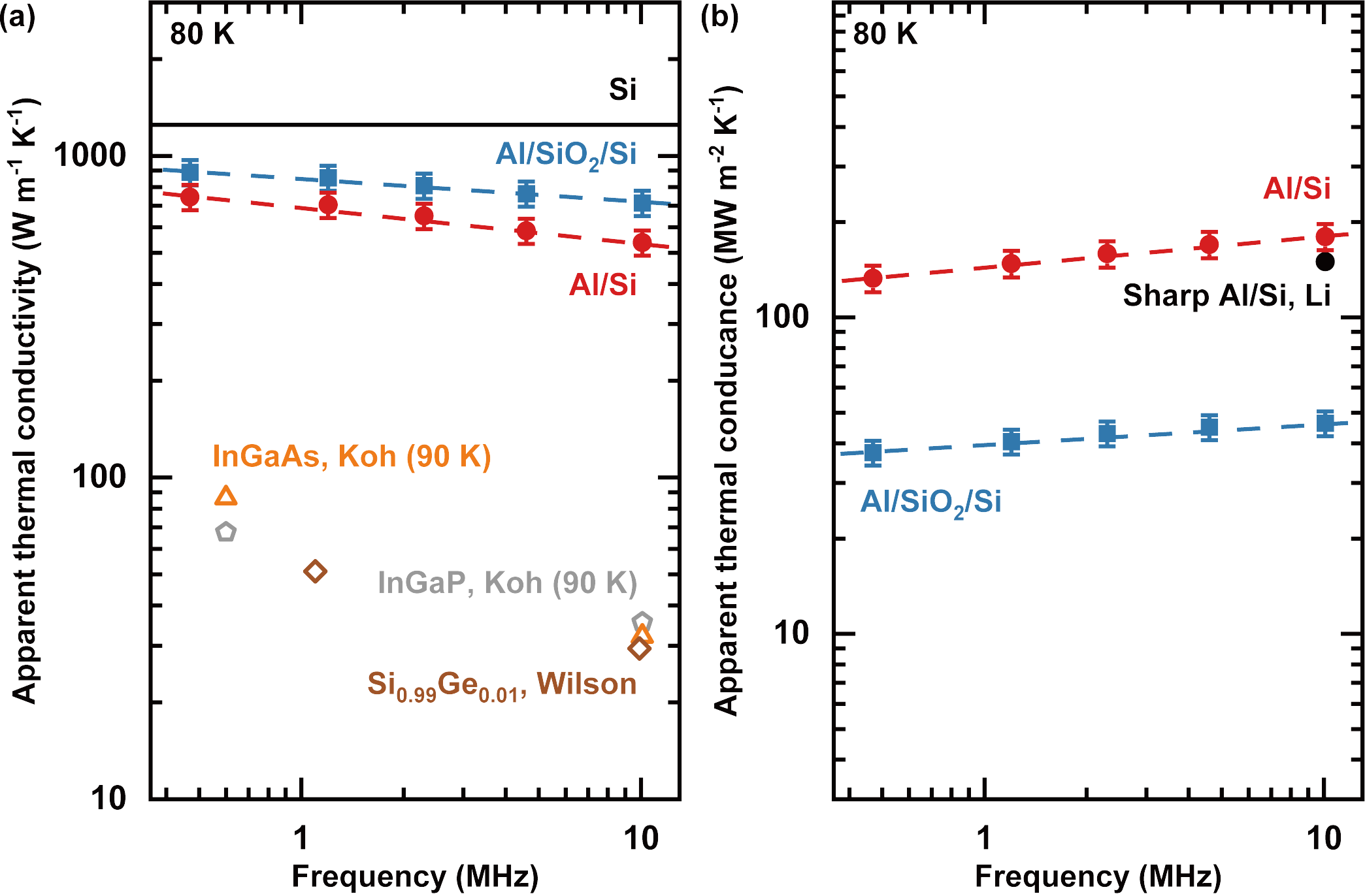}%
\caption{\label{fig_4} Apparent (a) $\Lambda$ and (b) \textit{G} of Al/Si (solid red circles) and Al/SiO$_2$/Si (solid blue squares) samples, derived from the one-channel thermal model at 80 K. The solid black line represents $\Lambda$ of bulk Si \cite{26} while the solid black circle stands for \textit{G} of sharp Al/Si interface by Li \textit{et al} \cite{20}. Open brown diamonds exhibit reported frequency dependence in Si$_{0.99}$Ge$_{0.01}$ \cite{18}. Orange triangles and grey pentagons show the frequency-dependenct $\Lambda$ measured in In$_{0.53}$Ga$_{0.47}$As and In$_{0.49}$Ga$_{0.51}$P, respectively \cite{10}.}
\end{figure}

\subsection{Cumulative thermal conductivity with phonon mean free path}

Frequency dependent $\Lambda$ has achieved great success in probing the spectral distribution of phonon MFPs in semiconductor alloys and 2D materials, where we assume phonons with MFPs longer than the characteristic length undergo ballistic transport and do not contribute to the measured $\Lambda$ \cite{10,30}. The characteristic length was parameterized as twice the thermal penetration depth \textit{d}=$\sqrt{\Lambda /\pi C f}$, where \textit{C} is the volumetric heat capacity. However, previous TDTR measurements on Si showed no frequency dependence, even though Si has a sufficiently broad spectrum of phonon MFPs. Although Regner \textit{et al.} reported a measured phonon MFPs spectrum in Si using BB-FDTR \cite{15}, their experiments have been questioned due to certain limitations, as extensively discussed before \cite{9,18}. In contrast, our study provides a more accurate approximation of the phonon MFP distribution in silicon at room temperature and 80 K, where we consider ballistic phonons with characteristic lengths larger than 2\textit{d} do not contribute thermal conductivity, as illustrated in Fig. 5. For the Al/Si sample, our data exhibit remarkable agreement with the first-principles calculation by Esfarjani \textit{et al.} \cite{19} at RT. At 80 K, our measurements align closely with first-principles results, calculated through interpolation of calculations by Minnich \textit{et al.} at 100 K and 60 K \cite{14}, with a slight deviation for phonons with short MFPs. This deviation is attributed to the mismatched phonon DOS between Al and Si, particularly for high-frequency phonons (higher than 10 THz) in Si, which are either reflected or decomposed to low-energy phonons at the interface. As a result, we can not accurately probe their contribution to the cumulative $\Lambda$. For the Al/SiO$_{2}$Si sample, the calculated cumulative $\Lambda$ with phonon MFP displays large derivations away from the first-principles calculation. The difference can be explained by the diffuse interface, which effectively reduces the ballistic phonon population, thereby limiting the extraction of intrinsic spectral phonon MFPs. 

This discovery emphasizes the feasibility of frequency dependence to accurately measure spectral phonon MFPs, depending on the proper selection of boundary conditions. To accurately probe the spectral phonon MFPs, a boundary condition with minimal phonon scattering at the transducer/sample interface is crucial. Specifically, not only an atomically sharp interface but also a metal transducer with a matched or larger Debye temperature is required. For example, we postulate that Al is a better transducer material, compared to Ta, to measure the spectral phonon MFPs of Si. There are two reasons. Firstly, Ta shows a more dissimilar Debye temperature to Si, leaving a higher interfacial phonon scattering rate, especially for high-energy phonons. Secondly, Ta has a higher density of states for low-energy phonons, leading to a smaller temperature difference between low- and high-energy phonons in Si, compared to the Al transducer.

On the contrary, if we want a TDTR measurement without any frequency dependence, a metal transducer with a low Debye temperature and a diffuse interface is preferred. In this sense, Ta or AuPd alloy is better than Al. This conclusion echoes the opinion of Wilson \textit{et al.} \cite{18}.
\begin{figure}
\includegraphics[width=0.48\textwidth,height=0.17\textwidth]{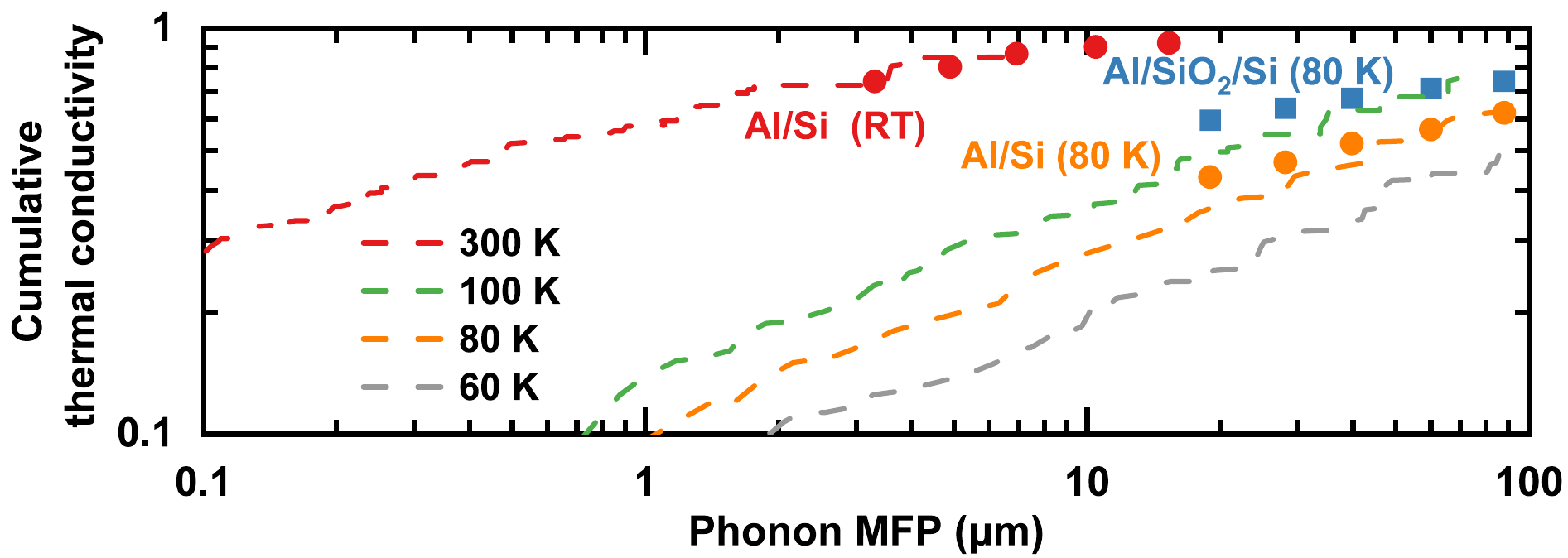}%
\caption{\label{fig_5} Phonon MFP spectra for Si at RT and 80 K. Dashed lines represent first-principles calculations of bulk Si at 300 K (red) by Esfarjani \textit{et al.} \cite{19}, 100 K (green), and 60 K (grey) \cite{14} by Minnich \textit{et al}. The dashed orange line denotes the cumulative $\Lambda$ at 80 K, calculated from interpolating values between 100 K and 60 K by Minnich \textit{et al}. Solid circles exhibit our measurements in the Al/Si sample at RT (red) and 80 K (orange), while solid blue squares display measurements on the Al/SiO$_{2}$/Si sample.}
\end{figure}

\subsection{Two-channel model for non-equilibrium thermal transport}
\label{subsec: Two-channel model for non-equilibrium thermal transport}

To quantitatively analyze the observed frequency dependence, we also applied the two-channel model to fit TDTR data  (see Supplemental Material \cite{25}). In the model, we divide the low-energy phonons (\textless 3 THz, Channel 1) and high-energy phonons (\textgreater 3 THz, Channel 2) into separate channels, where there is a coupling factor \textit{g} between the two channels. We select the partition frequency (3 THz) based on the principle that only phonon modes with linear dispersion and low heat capacity are considered in the low-frequency Channel 1 \cite{31}. The heat capacity of Channel 1 (\textit{C}$_{1}$) and Channel 2 (\textit{C}$_{2}$) are derived from calculations of the phonon density of states in Si \cite{31}. At RT, \textit{C}$_{1}$ = 3 $\times$ 10$^{4}$ J m$^{-3}$ K$^{-1}$ and \textit{C}$_{2}$ = 1.62 $\times$ 10$^{6}$ J m$^{-3}$ K$^{-1}$, while at 80 K, \textit{C}$_{1}$ = 8 $\times$ 10$^{3}$ J m$^{-3}$ K$^{-1}$ and \textit{C}$_{2}$ = 4.31 $\times$ 10$^{5}$ J m$^{-3}$ K$^{-1}$. Acquired TDTR signals can be well fit for the entire frequency range, using a set of thermal conductivity contributed by Channel 1 ($\Lambda_{1}$) and Channel 2 ($\Lambda_{2}$), interfacial thermal conductance from Channel 1 (\textit{G}$_{1}$) and Channel 2 (\textit{G}$_{2}$), and \textit{g}. We show the acquired total $\Lambda$ and \textit{G} in Fig. 6, with a comparison of previous calculations and experimental data. In the Al/Si sample, we obtained $\Lambda_{1}$ = 66 W m$^{-1}$ K$^{-1}$, $\Lambda_{2}$ = 75 W m$^{-1}$ K$^{-1}$, \textit{G}$_{1}$ = 58 MW m$^{-2}$ K$^{-1}$, \textit{G}$_{2}$ = 400 MW m$^{-2}$ K$^{-1}$, and \textit{g} = 1.2 $\times$ 10$^{14}$ W m$^{-3}$ K$^{-1}$ at RT. The total thermal conductivity $\Lambda_{1}$ + $\Lambda_{2}$ = 141 W m$^{-1}$ K$^{-1}$ aligns well with previously reported value \cite{18,26,32}. Low-frequency phonons in Channel 1 contribute 46.8 $\%$ of the total $\Lambda$ while high-frequency phonons contribute 53.2 $\%$, consistent with theoretical predictions \cite{33}. The total thermal conductance \textit{G}$_{1}$ + \textit{G}$_{2}$ = 458 MW m$^{-2}$ K$^{-1}$ matched the value predicted by the DMM \cite{20}, even without considering the serial interfacial thermal resistance from electron-phonon interactions within the Al layer \cite{34}. The result suggests the presence of inelastic phonon transport in the Al/Si interface at RT. The obtained coupling factor \textit{g} lies in the order of 10$^{14}$ W m$^{-3}$ K$^{-1}$, which is expected by the scattering rates of low-frequency phonons \cite{35,36}. Additionally, at 80 K, we obtained $\Lambda_{1}$ = 630 W m$^{-1}$ K$^{-1}$, $\Lambda_{2}$ = 320 W m$^{-1}$ K$^{-1}$, \textit{G}$_{1}$ = 22 MW m$^{-2}$ K$^{-1}$, \textit{G}$_{2}$ = 162 MW m$^{-2}$ K$^{-1}$, and \textit{g} = 5 $\times$ 10$^{12}$ W m$^{-3}$ K$^{-1}$. The total $\Lambda$ of 950 W m$^{-1}$ K$^{-1}$ is still lower than previously reported values in Si, which probably arises from the suppressed $\Lambda$$_{//}$ due to a not large enough  laser spot size. The partial $\Lambda$ contributed by Channel 1 increases to 66.5 $\%$, which is consistent with the expectation that low-energy phonons play a more significant part in $\Lambda$ at lower temperatures. Besides, the total \textit{G} = 184 MW m$^{-2}$ K$^{-1}$ confirms DMM predictions within the measurement uncertainty. Moreover, the coupling factor \textit{g} diminishes to the order of 10$^{12}$ W m$^{-3}$ K$^{-1}$, originating from the reduced phonon-phonon scattering rates at 80 K.
\begin{figure}
\includegraphics[width=0.48\textwidth,height=0.3\textwidth]{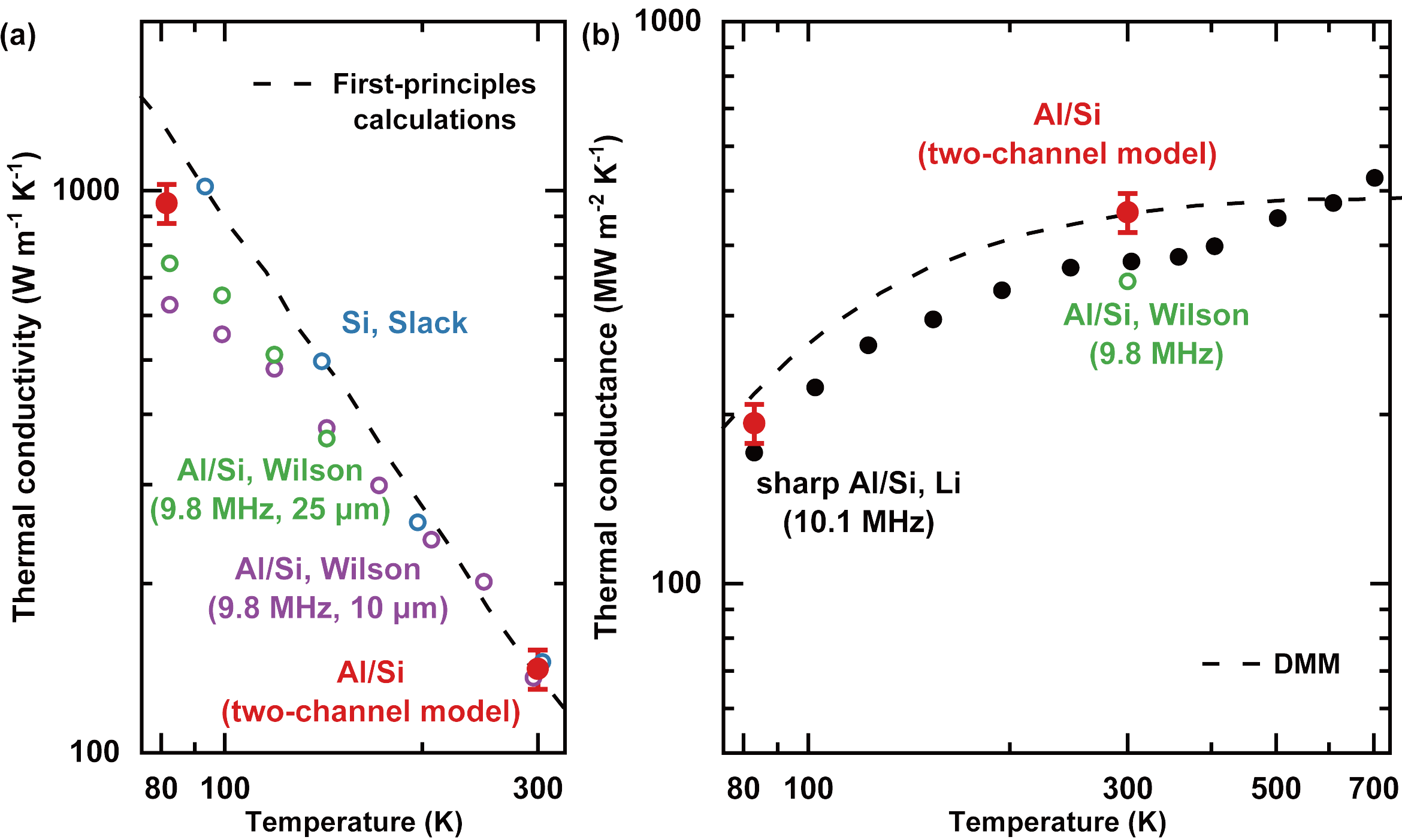}%
\caption{\label{fig_6}Total (a) $\Lambda$ and (b) \textit{G} of the Al/Si sample, extracted by the two-channel model (solid red circles). The dashed black lines represent predicted $\Lambda$ by first-principles calculations \cite{33} in (a) and \textit{G} by DMM \cite{20} in (b). Open circles in (a) denote $\Lambda$ of Si measured by Slack \textit{et al.} using steady-state techniques (blue) \cite{26} and by Wilson \textit{et al.} using TDTR, with a modulation frequency of 9.8 MHz and spot sizes of 25 $\mu$m (green) and 10 $\mu$m (purple) \cite{18}. Circles in (b) display \textit{G} of Al/Si interface measured by Wilson \textit{et al.} (9.8 MHz, green) \cite{18} and Li \textit{et al.} (10.1 MHz, black) \cite{20} using TDTR.} 
\end{figure}

\section{Summary}
\label{sec: Summary}

In summary, our TDTR measurements observed the frequency dependent thermal transport in Si with an atomically sharp Al/Si interface at RT. The dependence is previously buried due to the lack of high-quality sharp interfaces, as validated by measurements on the Al/SiO$_{2}$/Si sample. In this context, imperfections at the interface would facilitate energy exchange between different phonon branches and prohibit the non-equilibrium effects. Moreover, the governance of the diffuse interface is restricted at cryogenic temperature, where phonons with longer MFPs exhibiting weaker interactions would enhance the non-equilibrium heat flow. Our findings offer a deeper comprehension of frequency dependence in TDTR experiments, expanding the purview of non-Fourier thermal transport. We propose that frequency dependence is a common phenomenon in TDTR experiments mediated by boundary conditions, rather than a material-specific anomaly, which is vital for accurately measuring thermal properties through TDTR. Furthermore, this work holds great importance for future heat dissipation strategies of high-frequency electronics.

\begin{acknowledgments}
We acknowledge the funding support from the National
Natural Science Foundation of China (Nos. 52161145502 and 12004211), the Shenzhen Science and Technology Program
(Nos. RCYX20200714114643187 and WDZC20200821100123001), the Tsinghua Shenzhen International Graduate School (Nos. QD2021008N and JC2021008), and the Guangdong Special Support Program (No. 2023TQ07A273).
\end{acknowledgments}

\bibliography{reference}

\begin{thebibliography}{39}%
\makeatletter
\providecommand \@ifxundefined [1]{%
 \@ifx{#1\undefined}
}%
\providecommand \@ifnum [1]{%
 \ifnum #1\expandafter \@firstoftwo
 \else \expandafter \@secondoftwo
 \fi
}%
\providecommand \@ifx [1]{%
 \ifx #1\expandafter \@firstoftwo
 \else \expandafter \@secondoftwo
 \fi
}%
\providecommand \natexlab [1]{#1}%
\providecommand \enquote  [1]{``#1''}%
\providecommand \bibnamefont  [1]{#1}%
\providecommand \bibfnamefont [1]{#1}%
\providecommand \citenamefont [1]{#1}%
\providecommand \href@noop [0]{\@secondoftwo}%
\providecommand \href [0]{\begingroup \@sanitize@url \@href}%
\providecommand \@href[1]{\@@startlink{#1}\@@href}%
\providecommand \@@href[1]{\endgroup#1\@@endlink}%
\providecommand \@sanitize@url [0]{\catcode `\\12\catcode `\$12\catcode `\&12\catcode `\#12\catcode `\^12\catcode `\_12\catcode `\%12\relax}%
\providecommand \@@startlink[1]{}%
\providecommand \@@endlink[0]{}%
\providecommand \url  [0]{\begingroup\@sanitize@url \@url }%
\providecommand \@url [1]{\endgroup\@href {#1}{\urlprefix }}%
\providecommand \urlprefix  [0]{URL }%
\providecommand \Eprint [0]{\href }%
\providecommand \doibase [0]{https://doi.org/}%
\providecommand \selectlanguage [0]{\@gobble}%
\providecommand \bibinfo  [0]{\@secondoftwo}%
\providecommand \bibfield  [0]{\@secondoftwo}%
\providecommand \translation [1]{[#1]}%
\providecommand \BibitemOpen [0]{}%
\providecommand \bibitemStop [0]{}%
\providecommand \bibitemNoStop [0]{.\EOS\space}%
\providecommand \EOS [0]{\spacefactor3000\relax}%
\providecommand \BibitemShut  [1]{\csname bibitem#1\endcsname}%
\let\auto@bib@innerbib\@empty
\bibitem [{\citenamefont {Chen}(2021)}]{1}%
  \BibitemOpen
  \bibfield  {author} {\bibinfo {author} {\bibfnamefont {G.}~\bibnamefont {Chen}},\ }\bibfield  {title} {\bibinfo {title} {Non-fourier phonon heat conduction at the microscale and nanoscale},\ }\href@noop {} {\bibfield  {journal} {\bibinfo  {journal} {Nat. Rev. Phys.}\ }\textbf {\bibinfo {volume} {3}},\ \bibinfo {pages} {555} (\bibinfo {year} {2021})}\BibitemShut {NoStop}%
\bibitem [{\citenamefont {Mahan}\ and\ \citenamefont {Claro}(1988)}]{2}%
  \BibitemOpen
  \bibfield  {author} {\bibinfo {author} {\bibfnamefont {G.}~\bibnamefont {Mahan}}\ and\ \bibinfo {author} {\bibfnamefont {F.}~\bibnamefont {Claro}},\ }\bibfield  {title} {\bibinfo {title} {Nonlocal theory of thermal conductivity},\ }\href@noop {} {\bibfield  {journal} {\bibinfo  {journal} {Phys. Rev. B}\ }\textbf {\bibinfo {volume} {38}},\ \bibinfo {pages} {1963} (\bibinfo {year} {1988})}\BibitemShut {NoStop}%
\bibitem [{\citenamefont {Kanatzidis}(2010)}]{3}%
  \BibitemOpen
  \bibfield  {author} {\bibinfo {author} {\bibfnamefont {M.~G.}\ \bibnamefont {Kanatzidis}},\ }\bibfield  {title} {\bibinfo {title} {Nanostructured thermoelectrics: the new paradigm?},\ }\href@noop {} {\bibfield  {journal} {\bibinfo  {journal} {Chem. Mater.}\ }\textbf {\bibinfo {volume} {22}},\ \bibinfo {pages} {648} (\bibinfo {year} {2010})}\BibitemShut {NoStop}%
\bibitem [{\citenamefont {Pop}(2010)}]{4}%
  \BibitemOpen
  \bibfield  {author} {\bibinfo {author} {\bibfnamefont {E.}~\bibnamefont {Pop}},\ }\bibfield  {title} {\bibinfo {title} {Energy dissipation and transport in nanoscale devices},\ }\href@noop {} {\bibfield  {journal} {\bibinfo  {journal} {Nano Res.}\ }\textbf {\bibinfo {volume} {3}},\ \bibinfo {pages} {147} (\bibinfo {year} {2010})}\BibitemShut {NoStop}%
\bibitem [{\citenamefont {Sverdrup}\ \emph {et~al.}(2001)\citenamefont {Sverdrup}, \citenamefont {Sinha}, \citenamefont {Asheghi}, \citenamefont {Uma},\ and\ \citenamefont {Goodson}}]{5}%
  \BibitemOpen
  \bibfield  {author} {\bibinfo {author} {\bibfnamefont {P.}~\bibnamefont {Sverdrup}}, \bibinfo {author} {\bibfnamefont {S.}~\bibnamefont {Sinha}}, \bibinfo {author} {\bibfnamefont {M.}~\bibnamefont {Asheghi}}, \bibinfo {author} {\bibfnamefont {S.}~\bibnamefont {Uma}},\ and\ \bibinfo {author} {\bibfnamefont {K.}~\bibnamefont {Goodson}},\ }\bibfield  {title} {\bibinfo {title} {Measurement of ballistic phonon conduction near hotspots in silicon},\ }\href@noop {} {\bibfield  {journal} {\bibinfo  {journal} {Appl. Phys. Lett.}\ }\textbf {\bibinfo {volume} {78}},\ \bibinfo {pages} {3331} (\bibinfo {year} {2001})}\BibitemShut {NoStop}%
\bibitem [{\citenamefont {Chiu}\ \emph {et~al.}(2005)\citenamefont {Chiu}, \citenamefont {Deshpande}, \citenamefont {Postma}, \citenamefont {Lau}, \citenamefont {Miko}, \citenamefont {Forro},\ and\ \citenamefont {Bockrath}}]{6}%
  \BibitemOpen
  \bibfield  {author} {\bibinfo {author} {\bibfnamefont {H.~Y.}\ \bibnamefont {Chiu}}, \bibinfo {author} {\bibfnamefont {V.~V.}\ \bibnamefont {Deshpande}}, \bibinfo {author} {\bibfnamefont {H.~W.~C.}\ \bibnamefont {Postma}}, \bibinfo {author} {\bibfnamefont {C.~N.}\ \bibnamefont {Lau}}, \bibinfo {author} {\bibfnamefont {C.}~\bibnamefont {Miko}}, \bibinfo {author} {\bibfnamefont {L.}~\bibnamefont {Forro}},\ and\ \bibinfo {author} {\bibfnamefont {M.}~\bibnamefont {Bockrath}},\ }\bibfield  {title} {\bibinfo {title} {Ballistic phonon thermal transport in multiwalled carbon nanotubes},\ }\href@noop {} {\bibfield  {journal} {\bibinfo  {journal} {Phys. Rev. Lett.}\ }\textbf {\bibinfo {volume} {95}},\ \bibinfo {pages} {226101} (\bibinfo {year} {2005})}\BibitemShut {NoStop}%
\bibitem [{\citenamefont {Chen}(1998)}]{7}%
  \BibitemOpen
  \bibfield  {author} {\bibinfo {author} {\bibfnamefont {G.}~\bibnamefont {Chen}},\ }\bibfield  {title} {\bibinfo {title} {Thermal conductivity and ballistic-phonon transport in the cross-plane direction of superlattices},\ }\href@noop {} {\bibfield  {journal} {\bibinfo  {journal} {Phys. Rev. B}\ }\textbf {\bibinfo {volume} {57}},\ \bibinfo {pages} {14958} (\bibinfo {year} {1998})}\BibitemShut {NoStop}%
\bibitem [{\citenamefont {Chen}(2001)}]{8}%
  \BibitemOpen
  \bibfield  {author} {\bibinfo {author} {\bibfnamefont {G.}~\bibnamefont {Chen}},\ }\bibfield  {title} {\bibinfo {title} {Ballistic-diffusive heat-conduction equations},\ }\href@noop {} {\bibfield  {journal} {\bibinfo  {journal} {Phys. Rev. Lett.}\ }\textbf {\bibinfo {volume} {86}},\ \bibinfo {pages} {2297} (\bibinfo {year} {2001})}\BibitemShut {NoStop}%
\bibitem [{\citenamefont {Koh}\ \emph {et~al.}(2014)\citenamefont {Koh}, \citenamefont {Cahill},\ and\ \citenamefont {Sun}}]{9}%
  \BibitemOpen
  \bibfield  {author} {\bibinfo {author} {\bibfnamefont {Y.~K.}\ \bibnamefont {Koh}}, \bibinfo {author} {\bibfnamefont {D.~G.}\ \bibnamefont {Cahill}},\ and\ \bibinfo {author} {\bibfnamefont {B.}~\bibnamefont {Sun}},\ }\bibfield  {title} {\bibinfo {title} {Nonlocal theory for heat transport at high frequencies},\ }\href@noop {} {\bibfield  {journal} {\bibinfo  {journal} {Phys. Rev. B}\ }\textbf {\bibinfo {volume} {90}},\ \bibinfo {pages} {205412} (\bibinfo {year} {2014})}\BibitemShut {NoStop}%
\bibitem [{\citenamefont {Koh}\ and\ \citenamefont {Cahill}(2007)}]{10}%
  \BibitemOpen
  \bibfield  {author} {\bibinfo {author} {\bibfnamefont {Y.~K.}\ \bibnamefont {Koh}}\ and\ \bibinfo {author} {\bibfnamefont {D.~G.}\ \bibnamefont {Cahill}},\ }\bibfield  {title} {\bibinfo {title} {Frequency dependence of the thermal conductivity of semiconductor alloys},\ }\href@noop {} {\bibfield  {journal} {\bibinfo  {journal} {Phys. Rev. B}\ }\textbf {\bibinfo {volume} {76}},\ \bibinfo {pages} {075207} (\bibinfo {year} {2007})}\BibitemShut {NoStop}%
\bibitem [{\citenamefont {Sun}\ \emph {et~al.}(2024)\citenamefont {Sun}, \citenamefont {Zhao}, \citenamefont {Chen}, \citenamefont {Hu}, \citenamefont {Zhi}, \citenamefont {Wu}, \citenamefont {Kang}, \citenamefont {Tian},\ and\ \citenamefont {Gu}}]{11}%
  \BibitemOpen
  \bibfield  {author} {\bibinfo {author} {\bibfnamefont {B.}~\bibnamefont {Sun}}, \bibinfo {author} {\bibfnamefont {L.}~\bibnamefont {Zhao}}, \bibinfo {author} {\bibfnamefont {Z.}~\bibnamefont {Chen}}, \bibinfo {author} {\bibfnamefont {S.}~\bibnamefont {Hu}}, \bibinfo {author} {\bibfnamefont {A.}~\bibnamefont {Zhi}}, \bibinfo {author} {\bibfnamefont {J.}~\bibnamefont {Wu}}, \bibinfo {author} {\bibfnamefont {F.}~\bibnamefont {Kang}}, \bibinfo {author} {\bibfnamefont {X.}~\bibnamefont {Tian}},\ and\ \bibinfo {author} {\bibfnamefont {X.}~\bibnamefont {Gu}},\ }\href@noop {} {\bibinfo {title} {Ultrahigh through-plane thermal conductivity of graphite by reducing inter-plane twist}},\ \bibinfo {howpublished} {PREPRINT (Version 1) Research Square:10.21203/rs.3.rs-5186005/v1} (\bibinfo {year} {2024})\BibitemShut {NoStop}%
\bibitem [{\citenamefont {Sun}\ \emph {et~al.}(2017)\citenamefont {Sun}, \citenamefont {Gu}, \citenamefont {Zeng}, \citenamefont {Huang}, \citenamefont {Yan}, \citenamefont {Liu}, \citenamefont {Yang},\ and\ \citenamefont {Koh}}]{12}%
  \BibitemOpen
  \bibfield  {author} {\bibinfo {author} {\bibfnamefont {B.}~\bibnamefont {Sun}}, \bibinfo {author} {\bibfnamefont {X.}~\bibnamefont {Gu}}, \bibinfo {author} {\bibfnamefont {Q.}~\bibnamefont {Zeng}}, \bibinfo {author} {\bibfnamefont {X.}~\bibnamefont {Huang}}, \bibinfo {author} {\bibfnamefont {Y.}~\bibnamefont {Yan}}, \bibinfo {author} {\bibfnamefont {Z.}~\bibnamefont {Liu}}, \bibinfo {author} {\bibfnamefont {R.}~\bibnamefont {Yang}},\ and\ \bibinfo {author} {\bibfnamefont {Y.~K.}\ \bibnamefont {Koh}},\ }\bibfield  {title} {\bibinfo {title} {Temperature dependence of anisotropic thermal-conductivity tensor of bulk black phosphorus},\ }\href@noop {} {\bibfield  {journal} {\bibinfo  {journal} {Adv. Mater.}\ }\textbf {\bibinfo {volume} {29}},\ \bibinfo {pages} {1603297} (\bibinfo {year} {2017})}\BibitemShut {NoStop}%
\bibitem [{\citenamefont {Jiang}\ \emph {et~al.}(2017)\citenamefont {Jiang}, \citenamefont {Qian}, \citenamefont {Gu},\ and\ \citenamefont {Yang}}]{13}%
  \BibitemOpen
  \bibfield  {author} {\bibinfo {author} {\bibfnamefont {P.}~\bibnamefont {Jiang}}, \bibinfo {author} {\bibfnamefont {X.}~\bibnamefont {Qian}}, \bibinfo {author} {\bibfnamefont {X.}~\bibnamefont {Gu}},\ and\ \bibinfo {author} {\bibfnamefont {R.}~\bibnamefont {Yang}},\ }\bibfield  {title} {\bibinfo {title} {Probing anisotropic thermal conductivity of transition metal dichalcogenides {MX2 (M= Mo, W and X= S, Se)} using time-domain thermoreflectance},\ }\href@noop {} {\bibfield  {journal} {\bibinfo  {journal} {Adv. Mater.}\ }\textbf {\bibinfo {volume} {29}},\ \bibinfo {pages} {1701068} (\bibinfo {year} {2017})}\BibitemShut {NoStop}%
\bibitem [{\citenamefont {Minnich}\ \emph {et~al.}(2011)\citenamefont {Minnich}, \citenamefont {Johnson}, \citenamefont {Schmidt}, \citenamefont {Esfarjani}, \citenamefont {Dresselhaus}, \citenamefont {Nelson},\ and\ \citenamefont {Chen}}]{14}%
  \BibitemOpen
  \bibfield  {author} {\bibinfo {author} {\bibfnamefont {A.~J.}\ \bibnamefont {Minnich}}, \bibinfo {author} {\bibfnamefont {J.~A.}\ \bibnamefont {Johnson}}, \bibinfo {author} {\bibfnamefont {A.~J.}\ \bibnamefont {Schmidt}}, \bibinfo {author} {\bibfnamefont {K.}~\bibnamefont {Esfarjani}}, \bibinfo {author} {\bibfnamefont {M.~S.}\ \bibnamefont {Dresselhaus}}, \bibinfo {author} {\bibfnamefont {K.~A.}\ \bibnamefont {Nelson}},\ and\ \bibinfo {author} {\bibfnamefont {G.}~\bibnamefont {Chen}},\ }\bibfield  {title} {\bibinfo {title} {Thermal conductivity spectroscopy technique to measure phonon mean free paths},\ }\href@noop {} {\bibfield  {journal} {\bibinfo  {journal} {Phys. Rev. Lett.}\ }\textbf {\bibinfo {volume} {107}},\ \bibinfo {pages} {095901} (\bibinfo {year} {2011})}\BibitemShut {NoStop}%
\bibitem [{\citenamefont {Regner}\ \emph {et~al.}(2013)\citenamefont {Regner}, \citenamefont {Sellan}, \citenamefont {Su}, \citenamefont {Amon}, \citenamefont {McGaughey},\ and\ \citenamefont {Malen}}]{15}%
  \BibitemOpen
  \bibfield  {author} {\bibinfo {author} {\bibfnamefont {K.~T.}\ \bibnamefont {Regner}}, \bibinfo {author} {\bibfnamefont {D.~P.}\ \bibnamefont {Sellan}}, \bibinfo {author} {\bibfnamefont {Z.}~\bibnamefont {Su}}, \bibinfo {author} {\bibfnamefont {C.~H.}\ \bibnamefont {Amon}}, \bibinfo {author} {\bibfnamefont {A.~J.}\ \bibnamefont {McGaughey}},\ and\ \bibinfo {author} {\bibfnamefont {J.~A.}\ \bibnamefont {Malen}},\ }\bibfield  {title} {\bibinfo {title} {Broadband phonon mean free path contributions to thermal conductivity measured using frequency domain thermoreflectance},\ }\href@noop {} {\bibfield  {journal} {\bibinfo  {journal} {Nat. Commun.}\ }\textbf {\bibinfo {volume} {4}},\ \bibinfo {pages} {1640} (\bibinfo {year} {2013})}\BibitemShut {NoStop}%
\bibitem [{\citenamefont {Freedman}\ \emph {et~al.}(2013)\citenamefont {Freedman}, \citenamefont {Leach}, \citenamefont {Preble}, \citenamefont {Sitar}, \citenamefont {Davis},\ and\ \citenamefont {Malen}}]{16}%
  \BibitemOpen
  \bibfield  {author} {\bibinfo {author} {\bibfnamefont {J.~P.}\ \bibnamefont {Freedman}}, \bibinfo {author} {\bibfnamefont {J.~H.}\ \bibnamefont {Leach}}, \bibinfo {author} {\bibfnamefont {E.~A.}\ \bibnamefont {Preble}}, \bibinfo {author} {\bibfnamefont {Z.}~\bibnamefont {Sitar}}, \bibinfo {author} {\bibfnamefont {R.~F.}\ \bibnamefont {Davis}},\ and\ \bibinfo {author} {\bibfnamefont {J.~A.}\ \bibnamefont {Malen}},\ }\bibfield  {title} {\bibinfo {title} {Universal phonon mean free path spectra in crystalline semiconductors at high temperature},\ }\href@noop {} {\bibfield  {journal} {\bibinfo  {journal} {Sci. Rep.}\ }\textbf {\bibinfo {volume} {3}},\ \bibinfo {pages} {2963} (\bibinfo {year} {2013})}\BibitemShut {NoStop}%
\bibitem [{\citenamefont {Wilson}\ \emph {et~al.}(2013)\citenamefont {Wilson}, \citenamefont {Feser}, \citenamefont {Hohensee},\ and\ \citenamefont {Cahill}}]{17}%
  \BibitemOpen
  \bibfield  {author} {\bibinfo {author} {\bibfnamefont {R.}~\bibnamefont {Wilson}}, \bibinfo {author} {\bibfnamefont {J.~P.}\ \bibnamefont {Feser}}, \bibinfo {author} {\bibfnamefont {G.~T.}\ \bibnamefont {Hohensee}},\ and\ \bibinfo {author} {\bibfnamefont {D.~G.}\ \bibnamefont {Cahill}},\ }\bibfield  {title} {\bibinfo {title} {Two-channel model for nonequilibrium thermal transport in pump-probe experiments},\ }\href@noop {} {\bibfield  {journal} {\bibinfo  {journal} {Phys. Rev. B}\ }\textbf {\bibinfo {volume} {88}},\ \bibinfo {pages} {144305} (\bibinfo {year} {2013})}\BibitemShut {NoStop}%
\bibitem [{\citenamefont {Wilson}\ and\ \citenamefont {Cahill}(2014)}]{18}%
  \BibitemOpen
  \bibfield  {author} {\bibinfo {author} {\bibfnamefont {R.}~\bibnamefont {Wilson}}\ and\ \bibinfo {author} {\bibfnamefont {D.~G.}\ \bibnamefont {Cahill}},\ }\bibfield  {title} {\bibinfo {title} {Anisotropic failure of fourier theory in time-domain thermoreflectance experiments},\ }\href@noop {} {\bibfield  {journal} {\bibinfo  {journal} {Nat. Commun.}\ }\textbf {\bibinfo {volume} {5}},\ \bibinfo {pages} {5075} (\bibinfo {year} {2014})}\BibitemShut {NoStop}%
\bibitem [{\citenamefont {Esfarjani}\ \emph {et~al.}(2011)\citenamefont {Esfarjani}, \citenamefont {Chen},\ and\ \citenamefont {Stokes}}]{19}%
  \BibitemOpen
  \bibfield  {author} {\bibinfo {author} {\bibfnamefont {K.}~\bibnamefont {Esfarjani}}, \bibinfo {author} {\bibfnamefont {G.}~\bibnamefont {Chen}},\ and\ \bibinfo {author} {\bibfnamefont {H.~T.}\ \bibnamefont {Stokes}},\ }\bibfield  {title} {\bibinfo {title} {Heat transport in silicon from first-principles calculations},\ }\href@noop {} {\bibfield  {journal} {\bibinfo  {journal} {Phys. Rev. B}\ }\textbf {\bibinfo {volume} {84}},\ \bibinfo {pages} {085204} (\bibinfo {year} {2011})}\BibitemShut {NoStop}%
\bibitem [{\citenamefont {Li}\ \emph {et~al.}(2022)\citenamefont {Li}, \citenamefont {Liu}, \citenamefont {Hu}, \citenamefont {Song}, \citenamefont {Yang}, \citenamefont {Jiang}, \citenamefont {Wang}, \citenamefont {Koh}, \citenamefont {Zhao}, \citenamefont {Kang} \emph {et~al.}}]{20}%
  \BibitemOpen
  \bibfield  {author} {\bibinfo {author} {\bibfnamefont {Q.}~\bibnamefont {Li}}, \bibinfo {author} {\bibfnamefont {F.}~\bibnamefont {Liu}}, \bibinfo {author} {\bibfnamefont {S.}~\bibnamefont {Hu}}, \bibinfo {author} {\bibfnamefont {H.}~\bibnamefont {Song}}, \bibinfo {author} {\bibfnamefont {S.}~\bibnamefont {Yang}}, \bibinfo {author} {\bibfnamefont {H.}~\bibnamefont {Jiang}}, \bibinfo {author} {\bibfnamefont {T.}~\bibnamefont {Wang}}, \bibinfo {author} {\bibfnamefont {Y.~K.}\ \bibnamefont {Koh}}, \bibinfo {author} {\bibfnamefont {C.}~\bibnamefont {Zhao}}, \bibinfo {author} {\bibfnamefont {F.}~\bibnamefont {Kang}}, \emph {et~al.},\ }\bibfield  {title} {\bibinfo {title} {Inelastic phonon transport across atomically sharp metal/semiconductor interfaces},\ }\href@noop {} {\bibfield  {journal} {\bibinfo  {journal} {Nat. Commun.}\ }\textbf {\bibinfo {volume} {13}},\ \bibinfo {pages} {4901} (\bibinfo {year} {2022})}\BibitemShut {NoStop}%
\bibitem [{\citenamefont {Li}\ \emph {et~al.}(2023)\citenamefont {Li}, \citenamefont {Liu}, \citenamefont {Liu}, \citenamefont {Wang}, \citenamefont {Wang},\ and\ \citenamefont {Sun}}]{21}%
  \BibitemOpen
  \bibfield  {author} {\bibinfo {author} {\bibfnamefont {Q.}~\bibnamefont {Li}}, \bibinfo {author} {\bibfnamefont {F.}~\bibnamefont {Liu}}, \bibinfo {author} {\bibfnamefont {Y.}~\bibnamefont {Liu}}, \bibinfo {author} {\bibfnamefont {T.}~\bibnamefont {Wang}}, \bibinfo {author} {\bibfnamefont {X.}~\bibnamefont {Wang}},\ and\ \bibinfo {author} {\bibfnamefont {B.}~\bibnamefont {Sun}},\ }\bibfield  {title} {\bibinfo {title} {Effect of the alloyed interlayer on the thermal conductance of {Al/GaN} interface},\ }\href@noop {} {\bibfield  {journal} {\bibinfo  {journal} {J. Appl. Phys.}\ }\textbf {\bibinfo {volume} {134}} (\bibinfo {year} {2023})}\BibitemShut {NoStop}%
\bibitem [{\citenamefont {Cahill}(2004)}]{22}%
  \BibitemOpen
  \bibfield  {author} {\bibinfo {author} {\bibfnamefont {D.~G.}\ \bibnamefont {Cahill}},\ }\bibfield  {title} {\bibinfo {title} {Analysis of heat flow in layered structures for time-domain thermoreflectance},\ }\href@noop {} {\bibfield  {journal} {\bibinfo  {journal} {Rev. Sci. Instrum.}\ }\textbf {\bibinfo {volume} {75}},\ \bibinfo {pages} {5119} (\bibinfo {year} {2004})}\BibitemShut {NoStop}%
\bibitem [{\citenamefont {Sun}\ and\ \citenamefont {Koh}(2016)}]{23}%
  \BibitemOpen
  \bibfield  {author} {\bibinfo {author} {\bibfnamefont {B.}~\bibnamefont {Sun}}\ and\ \bibinfo {author} {\bibfnamefont {Y.~K.}\ \bibnamefont {Koh}},\ }\bibfield  {title} {\bibinfo {title} {Understanding and eliminating artifact signals from diffusely scattered pump beam in measurements of rough samples by time-domain thermoreflectance {(TDTR)}},\ }\href@noop {} {\bibfield  {journal} {\bibinfo  {journal} {Rev. Sci. Instrum.}\ }\textbf {\bibinfo {volume} {87}} (\bibinfo {year} {2016})}\BibitemShut {NoStop}%
\bibitem [{\citenamefont {Feldman}(1999)}]{24}%
  \BibitemOpen
  \bibfield  {author} {\bibinfo {author} {\bibfnamefont {A.}~\bibnamefont {Feldman}},\ }\bibfield  {title} {\bibinfo {title} {Algorithm for solutions of the thermal diffusion equation in a stratified medium with a modulated heating source},\ }\href@noop {} {\bibfield  {journal} {\bibinfo  {journal} {High Temp.-High Press.}\ }\textbf {\bibinfo {volume} {31}},\ \bibinfo {pages} {293} (\bibinfo {year} {1999})}\BibitemShut {NoStop}%
\bibitem [{25()}]{25}%
  \BibitemOpen
  \href@noop {} {\bibinfo  {journal} {See Supplementary Material at for detailed information on the one-channel model and the two-channel model for fitting TDTR measurements with sensitivity and uncertainty analysis}\ }\BibitemShut {NoStop}%
\bibitem [{\citenamefont {Glassbrenner}\ and\ \citenamefont {Slack}(1964)}]{26}%
  \BibitemOpen
\bibfield  {journal} {  }\bibfield  {author} {\bibinfo {author} {\bibfnamefont {C.~J.}\ \bibnamefont {Glassbrenner}}\ and\ \bibinfo {author} {\bibfnamefont {G.~A.}\ \bibnamefont {Slack}},\ }\bibfield  {title} {\bibinfo {title} {Thermal conductivity of silicon and germanium from 3 {K} to the melting point},\ }\href@noop {} {\bibfield  {journal} {\bibinfo  {journal} {Phys. Rev.}\ }\textbf {\bibinfo {volume} {134}},\ \bibinfo {pages} {A1058} (\bibinfo {year} {1964})}\BibitemShut {NoStop}%
\bibitem [{\citenamefont {Ding}\ \emph {et~al.}(2014)\citenamefont {Ding}, \citenamefont {Chen},\ and\ \citenamefont {Minnich}}]{27}%
  \BibitemOpen
  \bibfield  {author} {\bibinfo {author} {\bibfnamefont {D.}~\bibnamefont {Ding}}, \bibinfo {author} {\bibfnamefont {X.}~\bibnamefont {Chen}},\ and\ \bibinfo {author} {\bibfnamefont {A.}~\bibnamefont {Minnich}},\ }\bibfield  {title} {\bibinfo {title} {Radial quasiballistic transport in time-domain thermoreflectance studied using {Monte Carlo simulations}},\ }\href@noop {} {\bibfield  {journal} {\bibinfo  {journal} {Appl. Phys. Lett.}\ }\textbf {\bibinfo {volume} {104}} (\bibinfo {year} {2014})}\BibitemShut {NoStop}%
\bibitem [{\citenamefont {Gereth}\ and\ \citenamefont {Hubner}(1964)}]{28}%
  \BibitemOpen
  \bibfield  {author} {\bibinfo {author} {\bibfnamefont {R.}~\bibnamefont {Gereth}}\ and\ \bibinfo {author} {\bibfnamefont {K.}~\bibnamefont {Hubner}},\ }\bibfield  {title} {\bibinfo {title} {Phonon mean free path in silicon between 77 and 250 {K}},\ }\href@noop {} {\bibfield  {journal} {\bibinfo  {journal} {Phys. Rev.}\ }\textbf {\bibinfo {volume} {134}},\ \bibinfo {pages} {A235} (\bibinfo {year} {1964})}\BibitemShut {NoStop}%
\bibitem [{\citenamefont {Holland}(1964)}]{29}%
  \BibitemOpen
  \bibfield  {author} {\bibinfo {author} {\bibfnamefont {M.}~\bibnamefont {Holland}},\ }\bibfield  {title} {\bibinfo {title} {Phonon scattering in semiconductors from thermal conductivity studies},\ }\href@noop {} {\bibfield  {journal} {\bibinfo  {journal} {Phys. Rev.}\ }\textbf {\bibinfo {volume} {134}},\ \bibinfo {pages} {A471} (\bibinfo {year} {1964})}\BibitemShut {NoStop}%
\bibitem [{\citenamefont {Hu}\ \emph {et~al.}(2015)\citenamefont {Hu}, \citenamefont {Zeng}, \citenamefont {Minnich}, \citenamefont {Dresselhaus},\ and\ \citenamefont {Chen}}]{30}%
  \BibitemOpen
  \bibfield  {author} {\bibinfo {author} {\bibfnamefont {Y.}~\bibnamefont {Hu}}, \bibinfo {author} {\bibfnamefont {L.}~\bibnamefont {Zeng}}, \bibinfo {author} {\bibfnamefont {A.~J.}\ \bibnamefont {Minnich}}, \bibinfo {author} {\bibfnamefont {M.~S.}\ \bibnamefont {Dresselhaus}},\ and\ \bibinfo {author} {\bibfnamefont {G.}~\bibnamefont {Chen}},\ }\bibfield  {title} {\bibinfo {title} {Spectral mapping of thermal conductivity through nanoscale ballistic transport},\ }\href@noop {} {\bibfield  {journal} {\bibinfo  {journal} {Nat. Nanotechnol.}\ }\textbf {\bibinfo {volume} {10}},\ \bibinfo {pages} {701} (\bibinfo {year} {2015})}\BibitemShut {NoStop}%
\bibitem [{\citenamefont {Zdetsis}\ and\ \citenamefont {Wang}(1979)}]{31}%
  \BibitemOpen
  \bibfield  {author} {\bibinfo {author} {\bibfnamefont {A.}~\bibnamefont {Zdetsis}}\ and\ \bibinfo {author} {\bibfnamefont {C.}~\bibnamefont {Wang}},\ }\bibfield  {title} {\bibinfo {title} {Lattice dynamics of {Ge and Si} using the born-von karman model},\ }\href@noop {} {\bibfield  {journal} {\bibinfo  {journal} {Phys. Rev. B}\ }\textbf {\bibinfo {volume} {19}},\ \bibinfo {pages} {2999} (\bibinfo {year} {1979})}\BibitemShut {NoStop}%
\bibitem [{\citenamefont {Asheghi}\ \emph {et~al.}(2002)\citenamefont {Asheghi}, \citenamefont {Kurabayashi}, \citenamefont {Kasnavi},\ and\ \citenamefont {Goodson}}]{32}%
  \BibitemOpen
  \bibfield  {author} {\bibinfo {author} {\bibfnamefont {M.}~\bibnamefont {Asheghi}}, \bibinfo {author} {\bibfnamefont {K.}~\bibnamefont {Kurabayashi}}, \bibinfo {author} {\bibfnamefont {R.}~\bibnamefont {Kasnavi}},\ and\ \bibinfo {author} {\bibfnamefont {K.}~\bibnamefont {Goodson}},\ }\bibfield  {title} {\bibinfo {title} {Thermal conduction in doped single-crystal silicon films},\ }\href@noop {} {\bibfield  {journal} {\bibinfo  {journal} {J. Appl. Phys.}\ }\textbf {\bibinfo {volume} {91}},\ \bibinfo {pages} {5079} (\bibinfo {year} {2002})}\BibitemShut {NoStop}%
\bibitem [{\citenamefont {Wang}\ and\ \citenamefont {Huang}(2014)}]{33}%
  \BibitemOpen
  \bibfield  {author} {\bibinfo {author} {\bibfnamefont {X.}~\bibnamefont {Wang}}\ and\ \bibinfo {author} {\bibfnamefont {B.}~\bibnamefont {Huang}},\ }\bibfield  {title} {\bibinfo {title} {Computational study of in-plane phonon transport in {Si} thin films},\ }\href@noop {} {\bibfield  {journal} {\bibinfo  {journal} {Sci. Rep.}\ }\textbf {\bibinfo {volume} {4}},\ \bibinfo {pages} {6399} (\bibinfo {year} {2014})}\BibitemShut {NoStop}%
\bibitem [{\citenamefont {Majumdar}\ and\ \citenamefont {Reddy}(2004)}]{34}%
  \BibitemOpen
  \bibfield  {author} {\bibinfo {author} {\bibfnamefont {A.}~\bibnamefont {Majumdar}}\ and\ \bibinfo {author} {\bibfnamefont {P.}~\bibnamefont {Reddy}},\ }\bibfield  {title} {\bibinfo {title} {Role of electron-phonon coupling in thermal conductance of metal-nonmetal interfaces},\ }\href@noop {} {\bibfield  {journal} {\bibinfo  {journal} {Appl. Phys. Lett.}\ }\textbf {\bibinfo {volume} {84}},\ \bibinfo {pages} {4768} (\bibinfo {year} {2004})}\BibitemShut {NoStop}%
\bibitem [{\citenamefont {Ward}\ and\ \citenamefont {Broido}(2010)}]{35}%
  \BibitemOpen
  \bibfield  {author} {\bibinfo {author} {\bibfnamefont {A.}~\bibnamefont {Ward}}\ and\ \bibinfo {author} {\bibfnamefont {D.}~\bibnamefont {Broido}},\ }\bibfield  {title} {\bibinfo {title} {Intrinsic phonon relaxation times from first-principles studies of the thermal conductivities of {Si and Ge}},\ }\href@noop {} {\bibfield  {journal} {\bibinfo  {journal} {Phys. Rev. B}\ }\textbf {\bibinfo {volume} {81}},\ \bibinfo {pages} {085205} (\bibinfo {year} {2010})}\BibitemShut {NoStop}%
\bibitem [{\citenamefont {Maznev}\ \emph {et~al.}(2011)\citenamefont {Maznev}, \citenamefont {Johnson},\ and\ \citenamefont {Nelson}}]{36}%
  \BibitemOpen
  \bibfield  {author} {\bibinfo {author} {\bibfnamefont {A.~A.}\ \bibnamefont {Maznev}}, \bibinfo {author} {\bibfnamefont {J.~A.}\ \bibnamefont {Johnson}},\ and\ \bibinfo {author} {\bibfnamefont {K.~A.}\ \bibnamefont {Nelson}},\ }\bibfield  {title} {\bibinfo {title} {Onset of nondiffusive phonon transport in transient thermal grating decay},\ }\href@noop {} {\bibfield  {journal} {\bibinfo  {journal} {Phys. Rev. B}\ }\textbf {\bibinfo {volume} {84}},\ \bibinfo {pages} {195206} (\bibinfo {year} {2011})}\BibitemShut {NoStop}%
\bibitem [{\citenamefont {Sun}\ \emph {et~al.}(2020)\citenamefont {Sun}, \citenamefont {Niu}, \citenamefont {Hermann}, \citenamefont {Moon}, \citenamefont {Shulumba}, \citenamefont {Page}, \citenamefont {Zhao}, \citenamefont {Thind}, \citenamefont {Mahalingam}, \citenamefont {Milam-Guerrero} \emph {et~al.}}]{37}%
  \BibitemOpen
  \bibfield  {author} {\bibinfo {author} {\bibfnamefont {B.}~\bibnamefont {Sun}}, \bibinfo {author} {\bibfnamefont {S.}~\bibnamefont {Niu}}, \bibinfo {author} {\bibfnamefont {R.~P.}\ \bibnamefont {Hermann}}, \bibinfo {author} {\bibfnamefont {J.}~\bibnamefont {Moon}}, \bibinfo {author} {\bibfnamefont {N.}~\bibnamefont {Shulumba}}, \bibinfo {author} {\bibfnamefont {K.}~\bibnamefont {Page}}, \bibinfo {author} {\bibfnamefont {B.}~\bibnamefont {Zhao}}, \bibinfo {author} {\bibfnamefont {A.~S.}\ \bibnamefont {Thind}}, \bibinfo {author} {\bibfnamefont {K.}~\bibnamefont {Mahalingam}}, \bibinfo {author} {\bibfnamefont {J.}~\bibnamefont {Milam-Guerrero}}, \emph {et~al.},\ }\bibfield  {title} {\bibinfo {title} {High frequency atomic tunneling yields ultralow and glass-like thermal conductivity in chalcogenide single crystals},\ }\href@noop {} {\bibfield  {journal} {\bibinfo  {journal} {Nat. Commun.}\ }\textbf {\bibinfo {volume} {11}},\ \bibinfo {pages} {6039} (\bibinfo {year} {2020})}\BibitemShut {NoStop}%
\bibitem [{\citenamefont {Liu}\ \emph {et~al.}(2023)\citenamefont {Liu}, \citenamefont {Li}, \citenamefont {Qian}, \citenamefont {Yang}, \citenamefont {Wang}, \citenamefont {Li},\ and\ \citenamefont {Sun}}]{38}%
  \BibitemOpen
  \bibfield  {author} {\bibinfo {author} {\bibfnamefont {Y.}~\bibnamefont {Liu}}, \bibinfo {author} {\bibfnamefont {Q.}~\bibnamefont {Li}}, \bibinfo {author} {\bibfnamefont {Y.}~\bibnamefont {Qian}}, \bibinfo {author} {\bibfnamefont {Y.}~\bibnamefont {Yang}}, \bibinfo {author} {\bibfnamefont {S.}~\bibnamefont {Wang}}, \bibinfo {author} {\bibfnamefont {W.}~\bibnamefont {Li}},\ and\ \bibinfo {author} {\bibfnamefont {B.}~\bibnamefont {Sun}},\ }\bibfield  {title} {\bibinfo {title} {Thermal conductivity of high-temperature high-pressure synthesized {$\theta$-TaN}},\ }\href@noop {} {\bibfield  {journal} {\bibinfo  {journal} {Appl. Phys. Lett.}\ }\textbf {\bibinfo {volume} {122}} (\bibinfo {year} {2023})}\BibitemShut {NoStop}%
\bibitem [{\citenamefont {Wang}\ \emph {et~al.}(2024)\citenamefont {Wang}, \citenamefont {Luo}, \citenamefont {Chen}, \citenamefont {Zhou}, \citenamefont {Wang}, \citenamefont {Wu}, \citenamefont {Kang}, \citenamefont {Yu},\ and\ \citenamefont {Sun}}]{39}%
  \BibitemOpen
  \bibfield  {author} {\bibinfo {author} {\bibfnamefont {Y.}~\bibnamefont {Wang}}, \bibinfo {author} {\bibfnamefont {R.}~\bibnamefont {Luo}}, \bibinfo {author} {\bibfnamefont {J.}~\bibnamefont {Chen}}, \bibinfo {author} {\bibfnamefont {X.}~\bibnamefont {Zhou}}, \bibinfo {author} {\bibfnamefont {S.}~\bibnamefont {Wang}}, \bibinfo {author} {\bibfnamefont {J.}~\bibnamefont {Wu}}, \bibinfo {author} {\bibfnamefont {F.}~\bibnamefont {Kang}}, \bibinfo {author} {\bibfnamefont {K.}~\bibnamefont {Yu}},\ and\ \bibinfo {author} {\bibfnamefont {B.}~\bibnamefont {Sun}},\ }\bibfield  {title} {\bibinfo {title} {Proton collective quantum tunneling induces anomalous thermal conductivity of ice under pressure},\ }\href@noop {} {\bibfield  {journal} {\bibinfo  {journal} {Phys. Rev. Lett.}\ }\textbf {\bibinfo {volume} {132}},\ \bibinfo {pages} {264101} (\bibinfo {year} {2024})}\BibitemShut {NoStop}%
\end{thebibliography}%

\end{document}


\centerline{\textbf{\textit{Supplemental Material for}}}
\title{Boundary conditions dictate frequency dependence of thermal conductivity in silicon}

\author{Yizhe Liu}
\affiliation{Tsinghua SIGS, Tsinghua University, Shenzhen 518055, China}
\author{Qinshu Li}
\affiliation{Tsinghua-Berkeley Shenzhen Institute, Tsinghua University, Shenzhen 518055, China}
\author{Fang Liu}
\affiliation{State Key Laboratory for Mesoscopic Physics and Frontiers Science Center for Nano-optoelectronics, School of Physics, Peking University, Beijing 100871, China}
\author{Xinqiang Wang}
\affiliation{State Key Laboratory for Mesoscopic Physics and Frontiers Science Center for Nano-optoelectronics, School of Physics, Peking University, Beijing 100871, China}
\affiliation{Peking University Yangtze Delta Institute of Optoelectronics, Nantong, Jiangsu 226010, China}
\author{Bo Sun}
\email{sun.bo@sz.tsinghua.edu.cn}
\affiliation{Tsinghua SIGS, Tsinghua University, Shenzhen 518055, China}
\affiliation{Guangdong Provincial Key Laboratory of Thermal Management Engineering $\&$ Materials, Shenzhen 518055, China}

\maketitle
\clearpage

\section{\centerline{S1. One-channel model for conventional TDTR measurements}}
\subsection{1.1. TDTR setup}
We performed TDTR measurements to measure interfacial thermal conductance \textit{G} between the Al/Si interface and the thermal conductivity $\Lambda$ of Si. Briefly, a femtosecond laser (Coherent, Chameleon Ultra \uppercase\expandafter{\romannumeral2}) with a repetition rate of 80 MHz and a wavelength of 785 nm was applied. We employed high modulation frequencies from 10.1 MHz to 0.47 MHz for the pump beam through an electro-optic modulator (Conoptics, M350-210 for crystal, and 25D for driver). The probe beam was modulated at an audio frequency of 200 Hz by a mechanical chopper (Thorlabs, MC2000B). TDTR signals with respect to the delay time between pump and probe beams were demodulated through a lock-in amplifier (SRS, SR844). We carefully chose the laser powers (250 mW for the pump beam and 150 mW for the probe beam) to make sure that the steady-state temperature rises on the sample surface are less than 5 K for both room-temperature (RT) and cryogenic measurements.

\subsection{1.2. One-channel model}
For conventional TDTR, signals were analyzed by a diffusive multi-layer model, where thermal transport follows a one-channel Boltzmann transport equation (BTE) in cylindrical coordinates (\textit{r,z}). The governing equation of the one-channel model is based on Fourier's law, written as
\begin{equation}
C\frac{\partial T}{\partial t}=\frac{\Lambda_{//} }{r} \frac{\partial}{\partial t}(r\frac{\partial T}{\partial r})+\Lambda_{\perp} \frac{\partial^{2} T}{\partial z^{2}}
\label{eq:one}
\end{equation}
where \textit{C} is the volumetric heat capacity. This model holds the assumption that all energy carriers are in thermal equilibrium and transport diffusively for the whole process. Detailed solutions to the one-channel model have been extensively discussed by Feldman and Cahill \cite{22,24}.

\subsection{1.3. Sensitivity analysis}
\begin{figure}[H]
\centering
\includegraphics[width=1\textwidth,height=0.7\textwidth]{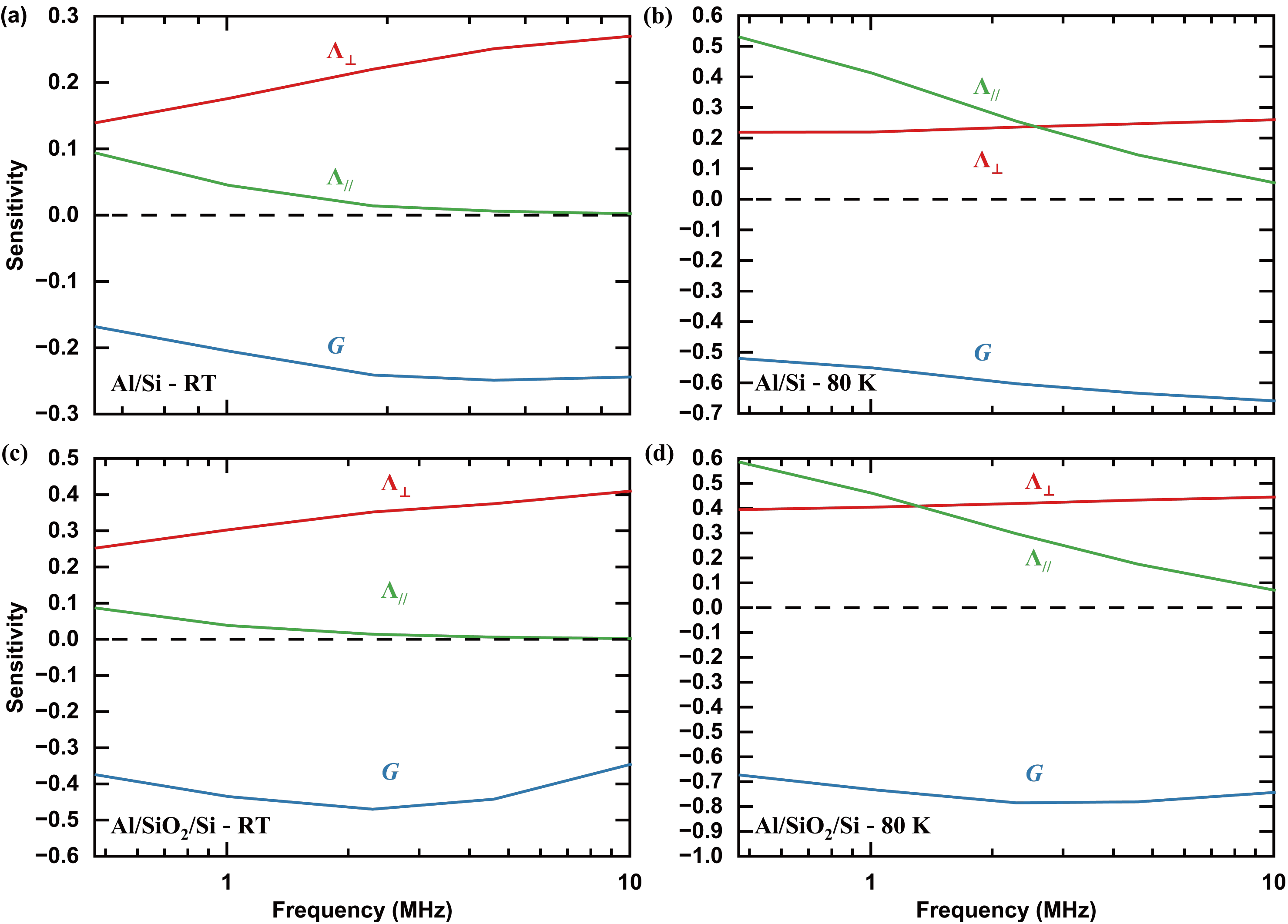}%
\caption{\label{fig_S1}Frequency-dependent sensitivities to \textit{G} (blue lines), $\Lambda$$_{\perp}$ (red lines), and $\Lambda$$_{//}$ (green lines) in the conventional one-channel model (at the delay time of 2 ns).}
\end{figure}

The reliability of using TDTR to measure an unknown thermal property hinges on the sensitivity of the measured signal to that specific quantity. Here, we calculate the sensitivity of thermal parameters in the one-channel model, which is defined as
\begin{equation}
S_{a}=\frac{\partial ln(\frac{-V_{in} }{V_{out}} ) }{\partial ln a},
\label{eq:two}
\end{equation}
where $\frac{-V_{in} }{V_{out}}$ is the TDTR signal, and \textit{a} is the parameter used for our thermal model. Fig. \ref{fig_S1} exhibits frequency-dependent sensitivities of \textit{G} and $\Lambda$ for the Al/Si sample and the Al/SiO$_{2}$/Si sample at RT and 80 K, respectively (at the delay time of 2 ns). High sensitivity values of \textit{G} and crossplane thermal conductivity $\Lambda$$_{\perp}$ span across the entire frequency range, indicating reliable TDTR measurements with minimal uncertainties. The sensitivity to inplane thermal conductivity $\Lambda$$_{//}$ is negligible for both samples at RT while exhibiting a notable value for modulation frequencies lower than 4 MHz at 80 K. The sensitivity analysis highlights the role of $\Lambda$$_{//}$ for low-frequency TDTR measurements at cryogenic temperatures, which would be suppressed due to the limited laser spot size, thus affecting the measured apparent $\Lambda$ \cite{14}.

\subsection{1.4. Uncertainty}
The TDTR measurement uncertainty of \textit{G} (or $\Lambda$) can be calculated using the following equation
\begin{equation}
\frac{\Delta G}{G}=\sqrt{\sum (\frac{S{_{a}} }{S_{G}}\frac{\Delta a}{a})^{2}+(\frac{S{_{\phi  }} }{S_{G}}\Delta \phi)^{2} }.
\label{eq:three}
\end{equation}
The sensitivity to parameters used in the one-channel model at a delay time of 2 ns is detailed in Table. \ref{table_S1}. Correspondingly, the calculated uncertainties for the parameters are presented in Table. \ref{table_S2}. The uncertainty of $\Lambda$ and \textit{G} is less than $\pm$4.5 $\%$ for both samples at RT. At 80 K, the uncertainty of \textit{G} shows a maximum value of $\pm$5.2 $\%$, while that of $\Lambda$ is less than $\pm$8.3 $\%$.
\clearpage

\section{\centerline{S2. Two-channel model for nonequilibrium thermal transport}}
\subsection{2.1. Two-channel model}
To quantify nonequilibrium heat flow between the low- and high-energy phonons in the Al/Si sample, we employed the two-channel model to fit acquired TDTR data. The governing BTEs in cylindrical coordinates (\textit{r,z}) are
\begin{equation}
C_{1}\frac{\partial T_{1}}{\partial t}=\frac{\Lambda_{1}}{r} \frac{\partial}{\partial t}(r\frac{\partial T_{1}}{\partial r})+\Lambda_{1}\frac{\partial^{2} T_{1}}{\partial z^{2}}+g(T_{2}-T_{1})
\label{eq:four}
\end{equation}
\begin{equation}
C_{l}\frac{\partial T_{2}}{\partial t}=\frac{\Lambda_{2}}{r} \frac{\partial}{\partial t}(r\frac{\partial T_{2}}{\partial r})+\Lambda_{2}\frac{\partial^{2} T_{2}}{\partial z^{2}}+g(T_{1}-T_{2})
\label{eq:five}
\end{equation}
where the subscript \textit{1} and \textit{2} denotes two channels for low-energy and high-energy phonons, respectively, and \textit{g} is the coupling factor between the two heat transfer channels. Detailed solutions to the two-channel model have been extensively discussed by Wilson \textit{et al}. \cite{17}
\subsection{2.2. TDTR data fitting}
\begin{figure}[H]
\centering
\includegraphics[width=1\textwidth,height=0.4\textwidth]{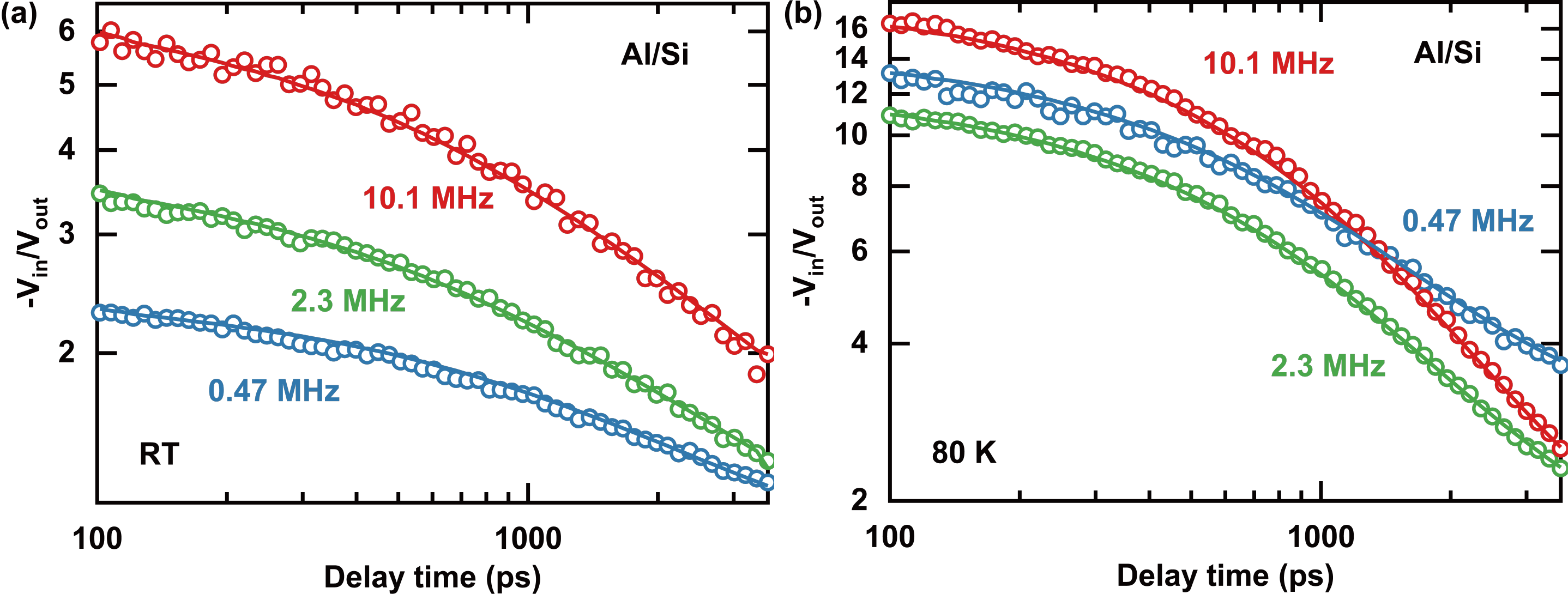}%
\caption{\label{fig_S2}Fits of the two-channel thermal model (solid lines) to TDTR signals (open circles) measured in the Al/Si sample at (a) RT and (b) 80 K. Red, green, and blue plots represent signals at frequencies of 10.1 MHz, 2.3 MHz, and 0.47 MHz, respectively.}
\end{figure}

Using the two-channel model, TDTR signals acquired from the Al/Si sample across the entire frequency range could be excellently fitted with one set of parameters ($\Lambda$$_{1}$, $\Lambda$$_{2}$, \textit{G}$_{1}$, \textit{G}$_{2}$, and \textit{g}), independent of modulation frequency. To extract multiple thermal properties, we iteratively fit the experimental signals for various frequencies according to the sensitivity of each parameter. The best-fitted parameters for the two-channel model are reported in the manuscript. The fitting curves from the two-channel model related to measured TDTR signals are depicted in Fig. \ref{fig_S2}.
\subsection{2.3. Sensitivity analysis}
\begin{figure}[H]
\centering
\includegraphics[width=1\textwidth,height=0.4\textwidth]{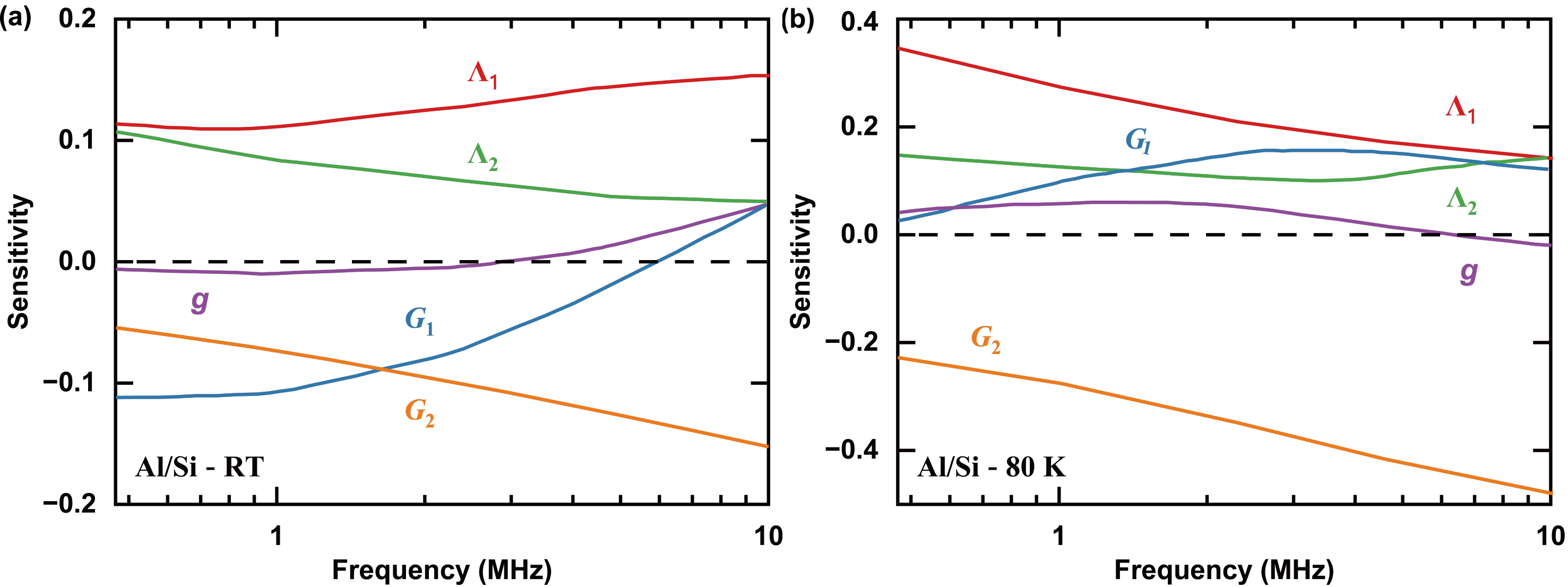}%
\caption{\label{fig_S3}Frequency-dependent sensitivities to $\Lambda$$_{1}$ (red), $\Lambda$$_{2}$ (green), \textit{G}$_{1}$ (blue), \textit{G}$_{2}$ (orange), and \textit{g} (purple) in the two-channel model (at the delay time of 2 ns).}
\end{figure}

The definition of sensitivity is given in Eq. S2. In the two-channel model, five thermophysical parameters are employed for the fitting process. Corresponding sensitivities to those parameters at the delay time of 2 ns are displayed in Fig. \ref{fig_S3}. At RT, $\Lambda$$_{1}$ exhibits pronounced sensitivity across the entire frequency range, whereas $\Lambda$$_{2}$ demonstrates sufficient sensitivity only at low frequencies. Conversely, \textit{G}$_{1}$ displays remarkable sensitivity at low frequencies while \textit{G}$_{2}$ gives a highest sensitivity at 10.1 MHz. When it comes to 80 K, $\Lambda$$_{1}$, $\Lambda$$_{2}$ and \textit{G}$_{2}$ show significant sensitivity for all frequencies, while \textit{G}$_{1}$ exhibitsonly at frequencies higher than 1 MHz. Notably, \textit{g} presents relatively low sensitivity, spanning all frequencies at both temperatures, suggesting a larger degree of uncertainty in our measurements.
\subsection{2.4. Uncertainty}
The definition of sensitivity is provided in Eq. S3. Given the frequency-dependent sensitivity of the fitting parameters, we selected the frequency point exhibiting the highest sensitivity for each parameter to calculate its corresponding uncertainty. The sensitivity of the parameters employed within the two-channel model at 2 ns delay time is delineated in Table S3. Correspondingly, the uncertainties for these parameters are presented in Table S4.
\clearpage

\clearpage
\clearpage
\clearpage

\begin{table}[H]
\centering
\caption{\label{table_S1}Sensitivity to parameters applied in the one-channel model}
\tabcolsep=0.15cm
\begin{tabular}{c  c  c  c  c  c  c  c  c}
\hline
\textbf{Al/Si at RT} & \textit{\textbf{S$_{h_{Al}}$}} & \textit{\textbf{S$_{\varLambda_{Al}}$}} & \textit{\textbf{S$_{C_{Al}}$}} & \textit{\textbf{S$_{C_{Si}}$}} & \textit{\textbf{S$_{w_{_{0}}}$}} & \textit{\textbf{S$_{\phi}$}} & \textit{\textbf{S$_{\varLambda_{Si}}$}} & \textit{\textbf{S$_{G}$}} \\ \hline
10.1 MHz&-0.210&-0.044&-0.256&+0.270&-0.004&+6.186&+0.270&-0.244\\
\hline
4.6 MHz&-0.162&-0.041&-0.206&+0.218&-0.009&+4.760&+0.251& -0.249\\
\hline
2.3 MHz&-0.127&-0.038&-0.167&+0.204&-0.019&+3.859&+0.220&-0.241\\
\hline
1.01 MHz&-0.088&-0.033&-0.123&+0.133&-0.046&+3.183&+0.176& -0.205\\
\hline
0.47 MHz&-0.066&-0.012&-0.095&+0.043&-0.103&+2.875&+0.139& -0.168\\
\hline
\end{tabular}
\\ \hspace*{\fill} \\
\begin{tabular}{c  c  c  c  c  c  c  c  c}
\hline
\textbf{Al/Si at 80 K} & \textit{\textbf{S$_{h_{Al}}$}} & \textit{\textbf{S$_{\varLambda_{Al}}$}} & \textit{\textbf{S$_{C_{Al}}$}} & \textit{\textbf{S$_{C_{Si}}$}} & \textit{\textbf{S$_{w_{_{0}}}$}} & \textit{\textbf{S$_{\phi}$}} & \textit{\textbf{S$_{\varLambda_{Si}}$}} & \textit{\textbf{S$_{G}$}} \\ \hline
10.1 MHz&+0.240&-0.051&+0.186&+0.201&-0.061&+17.059&+0.260&-0.659\\
\hline
4.6 MHz&+0.231&-0.047&+0.183&+0.073&-0.141&+12.182&+0.247& -0.634\\
\hline
2.3 MHz&+0.213&-0.042&+0.169&-0.021&-0.259&+11.289&+0.249&-0.603\\
\hline
1.01 MHz&+0.183&-0.036&+0.145&-0.176&-0.399&+11.488&+0.309& -0.551\\
\hline
0.47 MHz&+0.145&-0.034&+0.112&-0.280&-0.501&+13.375&+0.359& -0.520\\
\hline
\end{tabular}
\\ \hspace*{\fill} \\
\begin{tabular}{c  c  c  c  c  c  c  c  c}
\hline
\textbf{Al/SiO$_{2}$/Si at RT} & \textit{\textbf{S$_{h_{Al}}$}} & \textit{\textbf{S$_{\varLambda_{Al}}$}} & \textit{\textbf{S$_{C_{Al}}$}} & \textit{\textbf{S$_{C_{Si}}$}} & \textit{\textbf{S$_{w_{_{0}}}$}} & \textit{\textbf{S$_{\phi}$}} & \textit{\textbf{S$_{\varLambda_{Si}}$}} & \textit{\textbf{S$_{G}$}} \\ \hline
10.1 MHz&-0.373&-0.014&-0.388&+0.396&-0.004&+10.694&+0.410&-0.345\\
\hline
4.6 MHz&-0.297&-0.014&-0.313&+0.379&-0.008&+7.545&+0.386&-0.452\\
\hline
2.3 MHz&-0.211&-0.013&-0.226&+0.337&-0.017&+5.992&+0.352&-0.470\\
\hline
1.01 MHz&-0.148&-0.012&-0.163&+0.264&-0.040&+4.646&+0.303& -0.435\\
\hline
0.47 MHz&-0.110&-0.011&-0.123&+0.161&-0.096&+3.869&+0.252& -0.374\\
\hline
\end{tabular}
\\ \hspace*{\fill} \\
\begin{tabular}{c  c  c  c  c  c  c  c  c}
\hline
\textbf{Al/SiO$_{2}$/Si at 80 K} & \textit{\textbf{S$_{h_{Al}}$}} & \textit{\textbf{S$_{\varLambda_{Al}}$}} & \textit{\textbf{S$_{C_{Al}}$}} & \textit{\textbf{S$_{C_{Si}}$}} & \textit{\textbf{S$_{w_{_{0}}}$}} & \textit{\textbf{S$_{\phi}$}} & \textit{\textbf{S$_{\varLambda_{Si}}$}} & \textit{\textbf{S$_{G}$}} \\ \hline
10.1 MHz&-0.129&-0.015&-0.145&+0.376&-0.071&+34.929&+0.445&-0.743\\
\hline
4.6 MHz&-0.063&-0.015&-0.079&+0.264&-0.167&+26.937&+0.433& -0.781\\
\hline
2.3 MHz&-0.055&-0.014&-0.070&+0.124&-0.297&+24.041&+0.419&-0.793\\
\hline
1.01 MHz&-0.046&-0.013&-0.079&-0.055&-0.462&+24.441&+0.404& -0.722\\
\hline
0.47 MHz&-0.099&-0.011&-0.110&-0.199&-0.583&+28.935&+0.394& -0.673\\
\hline
\end{tabular}
\end{table}
\clearpage

\begin{table}[H]
\centering
\caption{\label{table_S2}Uncertainties for parameters applied in the one-channel model}
\tabcolsep=0.15cm
\begin{tabular}{c  c  c  c  c  c  c}
\hline
\textit{\textbf{}} & \textit{\textbf{h$_{Al}$}} & \textit{\textbf{$\varLambda$$_{Al}$}} & \textit{\textbf{C$_{Al}$}} & \textit{\textbf{C$_{Si}$}} & \textit{\textbf{w$_{0}$}} & \textit{\textbf{$\Delta$$\phi$}}\\
\hline
\textbf{Uncertainty}&$\pm$1$\%$&$\pm$10$\%$&$\pm$2$\%$&$\pm$2$\%$&$\pm$5$\%$& 0.001\\
\hline
\end{tabular}
\\ \hspace*{\fill} \\
\begin{tabular}{c  c  c  c  c  c  c}
\hline
\textbf{Uncertainty for $\Lambda$} & \textbf{10.1 MHz} & \textbf{4.6 MHz} & \textbf{2.3 MHz} & \textbf{1.01 MHz} & \textbf{0.47 MHz}\\
\hline
Al/Si at RT&$\pm$3.5$\%$&$\pm$3.1$\%$&$\pm$3.2$\%$&$\pm$3.3$\%$&$\pm$4.5$\%$\\
\hline
Al/Si at 80 K&$\pm$4.6$\%$&$\pm$4.6$\%$&$\pm$6.4$\%$&$\pm$6.9$\%$&$\pm$7.5$\%$\\
\hline
Al/SiO$_{2}$/Si at RT&$\pm$3.5$\%$&$\pm$2.9$\%$&$\pm$2.6$\%$&$\pm$2.4$\%$&$\pm$2.7$\%$\\
\hline
Al/SiO$_{2}$/Si at 80 K&$\pm$4.4$\%$&$\pm$3.9$\%$&$\pm$4.6$\%$&$\pm$6.5$\%$&$\pm$8.3$\%$\\
\hline
\end{tabular}
\\ \hspace*{\fill} \\
\begin{tabular}{c  c  c  c  c  c  c}
\hline
\textbf{Uncertainty for \textit{G}} & \textbf{10.1 MHz} & \textbf{4.6 MHz} & \textbf{2.3 MHz} & \textbf{1.01 MHz} & \textbf{0.47 MHz}\\
\hline
Al/Si at RT&$\pm$4.5$\%$&$\pm$3.1$\%$&$\pm$2.9$\%$&$\pm$2.8$\%$&$\pm$3.7$\%$\\
\hline
Al/Si at 80 K&$\pm$1.8$\%$&$\pm$1.8$\%$&$\pm$2.5$\%$&$\pm$3.9$\%$&$\pm$5.2$\%$\\
\hline
Al/SiO$_{2}$/Si at RT&$\pm$4.2$\%$&$\pm$2.4$\%$&$\pm$1.3$\%$&$\pm$1.7$\%$&$\pm$1.8$\%$\\
\hline
Al/SiO$_{2}$/Si at 80 K&$\pm$2.6$\%$&$\pm$2.2$\%$&$\pm$2.4$\%$&$\pm$4.4$\%$&$\pm$4.9$\%$\\
\hline
\end{tabular}
\end{table}
\clearpage

\begin{table}[H]
\centering
\caption{\label{table_S3}Sensitivity to parameters applied in the two-channel model}
\tabcolsep=0.15cm
\begin{tabular}{c  c  c  c  c  c  c  c  c}
\hline
\textbf{Al/Si at RT} &
\textit{\textbf{S$_{\varLambda_{Al}}$}} & 
\textit{\textbf{S$_{h_{Al}}$}} & 
\textit{\textbf{S$_{C_{Al}}$}} &
\textit{\textbf{S$_{C_{Si,1}}$}} & 
\textit{\textbf{S$_{C_{Si,2}}$}}&
\textit{\textbf{S$_{w_{_{0}}}$}} &
\textit{\textbf{S$_{\phi}$}} \\ \hline
10.1 MHz&-0.034&-0.243&-0.281&-0.003&+0.166&-0.011&+6.186\\
\hline
4.6 MHz&-0.031&-0.180&-0.217&+0.002&+0.166&-0.022&+4.760\\
\hline
2.3 MHz&-0.027&-0.131&-0.164&-0.001&+0.150&-0.043&+3.859\\
\hline
1.01 MHz&-0.020&-0.087&-0.116&-0.001&+0.099&-0.099&+3.183\\
\hline
0.47 MHz&-0.013&-0.060&-0.084&-0.002&+0.017&-0.210&+2.875\\
\hline
\end{tabular}
\\ \hspace*{\fill} \\
\begin{tabular}{c  c  c  c  c  c  c  c  c}
\hline
\textbf{Al/Si at 80 K} &\textit{\textbf{S$_{\varLambda_{Al}}$}} & 
\textit{\textbf{S$_{h_{Al}}$}} & 
\textit{\textbf{S$_{C_{Al}}$}} &
\textit{\textbf{S$_{C_{Si,1}}$}} & 
\textit{\textbf{S$_{C_{Si,2}}$}}&
\textit{\textbf{S$_{w_{_{0}}}$}} &
\textit{\textbf{S$_{\phi}$}} \\ \hline
10.1 MHz&-0.044&+0.225&+0.178&-0.020&+0.114&-0.185&+17.059\\
\hline
4.6 MHz&-0.039&+0.213&+0.170&-0.017&-0.021&-0.297&+12.182\\
\hline
2.3 MHz&-0.034&+0.192&+0.155&-0.014&-0.163&-0.450&+11.289\\
\hline
1.01 MHz&-0.029&+0.171&+0.139&-0.012&-0.304&-0.668&+11.488\\
\hline
0.47 MHz&-0.026&+0.158&+0.130&-0.011&-0.393&-0.861&+13.375\\
\hline
\end{tabular}
\end{table}
\clearpage

\begin{table}[H]
\centering
\caption{\label{table_S4}Uncertainties for parameters applied in the two-channel model}
\tabcolsep=0.15cm
\begin{tabular}{c  c  c  c  c  c  c  c}
\hline
\textit{\textbf{}} & \textit{\textbf{h$_{Al}$}} & \textit{\textbf{$\varLambda$$_{Al}$}} & \textit{\textbf{C$_{Al}$}} & \textit{\textbf{C$_{Si,1}$}}& \textit{\textbf{C$_{Si,2}$}} & \textit{\textbf{w$_{0}$}} & \textit{\textbf{$\Delta$$\phi$}}\\ \hline
\textbf{Uncertainty}&$\pm$1$\%$&$\pm$10$\%$&$\pm$2$\%$&$\pm$2$\%$&$\pm$2$\%$&$\pm$5$\%$& 0.001\\
\hline
\end{tabular}
\\ \hspace*{\fill} \\
\begin{tabular}{c  c  c  c  c  c}
\hline
\textbf{Uncertainty} &
\textit{\textbf{$\varLambda$$_{1}$}} &
\textit{\textbf{$\varLambda$$_{2}$}} &
\textit{\textbf{G$_{1}$}}& 
\textit{\textbf{G$_{2}$}}&
\textit{\textbf{g}}\\ \hline
\textbf{Al/Si at
RT}&$\pm$5.4$\%$&$\pm$10.1$\%$&$\pm$9.7$\%$&$\pm$5.5$\%$&$\pm$28.8$\%$\\
\hline
\textbf{Al/Si at
80 K}&$\pm$12.8$\%$&$\pm$9.2$\%$&$\pm$11.1$\%$&$\pm$3.2$\%$&$\pm$47.5$\%$\\
\hline
\end{tabular}
\end{table}
\clearpage
\bibliography{reference}